\documentclass[english,aps,twocolumn]{revtex4}
\usepackage[T1]{fontenc}
\usepackage[latin9]{inputenc}
\usepackage{graphicx}
\usepackage{amssymb}

\makeatletter

\makeatletter

\usepackage{natbib} 
\bibpunct{[}{]}{,}{s}{}{}

\makeatletter

\makeatletter

\makeatletter

\makeatletter

\makeatletter

\makeatletter

\usepackage{soul}

\usepackage{xcolor}

\makeatletter

\providecolor{lyxadded}{rgb}{0,0,1}
\providecolor{lyxdeleted}{rgb}{1,0,0}

\makeatletter

\usepackage{esint}



\makeatother

\makeatother

\makeatother

\makeatother

\makeatother

\makeatother

\makeatother

\makeatother

\makeatother

\usepackage{babel}
\makeatother

\begin{document}

\preprint{This line only printed with preprint option}

\title{A Micromagnetic Study of Magnetization Reversal in Ferromagnetic
Nanorings.}

\author{Gabriel D. Chaves-O'Flynn}

\email{gdc229@nyu.edu}

\affiliation{Department of Physics, New York University, 4 Washington Place, New
York, New York 10003, USA}

\author{A.D. Kent}

\email{andy.kent@nyu.edu}

\affiliation{Department of Physics, New York University, 4 Washington Place, New
York, New York 10003, USA}

\author{D.L. Stein}

\email{daniel.stein@nyu.edu}

\affiliation{Department of Physics, New York University, 4 Washington Place, New
York, New York 10003, USA}

\begin{abstract}
We present results of micromagnetic simulations of thin ferromagnetic
rings undergoing magnetization reversal. This geometry is one of few
examples in micromagnetics in which the transition states have been
found analytically in a $1D$ model. According to this model, at low
fields and large ring sizes, the energetically preferred transition
state is a localized magnetization fluctuation (instanton saddle).
At high fields and small ring size, the preferred saddle state is
a uniformly rotated magnetization (constant saddle). In the first
part of this paper, we use numerical micromagnetic simulations to
test these predictions of the $1D$ analytical model for more realistic
situations, including a variety of ring radii, annular widths and
magnetic fields. The predicted activation energies for magnetization
reversal are found to be in close agreement with numerical results,
even for rings with a large annular width where the $1D$ approximation
would be expected to break down. We find that this approximation breaks
down only when the ring's annular width exceeds its radius. In the
second part, we present new metastable states found in the large radius
limit and discuss how they provide a more complete understanding of
the energy landscape of magnetic nanorings. 
\end{abstract}
\maketitle

\section{Introduction}

\label{sec:intro}

The magnetic properties of thin ferromagnetic annuli have attracted
attention due to their potential applications in magnetic random access
memory: the absence in such geometries of edges or corners to nucleate
magnetization reversal leads to greater stability against reversal
than in other simply connected thin film elements.

There are several ways in which ferromagnetic annuli may be used as
memory elements, differing in the (meta)stable magnetization configurations
that represent a single bit. In all of these the magnetization lies
completely within the plane and its configuration is smooth everywhere.
One such pair of configurations is of opposite chirality, i.e., clockwise
or counterclockwise circulation of the magnetization \citep{_patterned_2007,zhu2000,yang2007};
another is the so-called {}``onion states\citep{castano2003,castano2006,rothman2001,Klaui2003,klaui2004,Laufenberg},
where there is a net total magnetization along a direction in the
ring's plane. In the former (latter) case, a circumferentially (uniformly)
directed magnetic field can be used to set the magnetization configuration.
The minimum energy configurations depend on the strength and direction
of the magnetic field as well as the relative dimensions of the ring
with respect to the exchange length.

Few analytic solutions have been obtained for the rate of thermally
induced reversal in micromagnetic problems, even in relatively simple
geometries. For thin annuli under the influence of a circumferential
magnetic field, however, the lowest energy transition (or saddle)
states have been found analytically~\citep{martens1} in a one-dimensional
approximation and studied numerically in the full three-dimensional
problem~\citep{chaves2008}.

The Kramers theory of reaction rates~\citep{Kramers40} can be used
to compute the typical lifetime of a given state when there are several
minimum energy configurations. The general form for the rate $\Gamma$
of thermally induced transitions between two minima in the limit of
low noise is given by the well-known Arrhenius formula~\citep{Hangii90}
$\Gamma=\Gamma_{0}\exp(-\Delta E/k_{B}T)$, where the prefactor $\Gamma_{0}$
is usually independent of the noise strength and depends only on the
shape of the energy landscape close to the extremal states relevant
to the transition. The activation energy, $\Delta E$, equals the
energy difference between the transition state and the metastable
state, thereby determining the stability of the latter. This is an
important figure of merit for memory devices.

In this paper, numerical micromagnetics are used to test the predictions
of the analytical theory of Martens et al.\textit{\/}~\citep{martens1}
for thermally induced transitions between states of opposite chirality
in a $1D$ approximation to the annular ring. The simulations were
made for a variety of mean radii, annular widths and magnetic fields.
\begin{figure}
\includegraphics[width=2.25in]{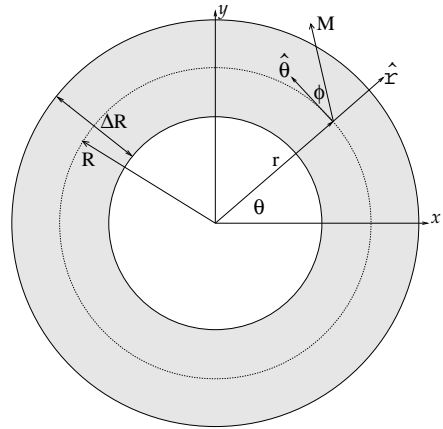}

\caption{Ring geometry showing the coordinates. The current runs along the
axis of the ring out of the page.}

\label{fig:geometry} 
\end{figure}

The geometry under study and accompanying relevant parameters are
represented in Fig.~\ref{fig:geometry}. The magnetic material is
in the shape of an annulus of mean radius~$R$, annular width~$\Delta R$
and (in the third dimension) thickness $t$. A current $I$ running
along the axis of the ring produces a circumferential external field
$\mathbf{H}(r)=(I/2\pi r)\hat{\theta}$. The ring is composed of a
soft isotropic ferromagnet (e.g., permalloy) with saturation magnetization
$M_{s}$ and exchange length $\lambda$. In all of the geometries
considered, the aspect ratio $k=t/R\ll1$, giving rise to magnetostatic
forces that constrain the magnetization to lie in the plane of the
ring ($M_{z}=0$)~\citep{martens1}. A magnetization configuration
can therefore be completely described by $\phi(\theta,r)$, the angle
the magnetization at a given radius makes with the unit vector lying
along the tangent to the circle with that radius: $\mathrm{\mathbf{M}}(\theta,r)=(M_{x},M_{y},M_{z})=M_{s}(\sin(\phi-\theta),\cos(\phi-\theta),0)$.

\section{Model}

\label{sec:model}

Our starting point is the Landau-Lifshitz-Gilbert~(LLG) equation~\citep{Landau35,Gilbert55}
\begin{equation}
\frac{d\mathbf{M}}{dt}=-|\gamma|\mathbf{M\times H_{\mathrm{eff}}-\frac{|\gamma|\alpha}{M_{s}}M\times(M\times H_{\mathrm{eff}})},\label{eq:LLG}\end{equation}
 where $\gamma$ is the gyromagnetic ratio and $\alpha$ is the (phenomenological)
damping constant. The effective magnetic field, $\mathbf{H_{\mathrm{eff}}}=-\nabla_{\mathbf{M}}E$,
contains all (external and internal) fields and is the variational
derivative of the total micromagnetic energy \begin{equation}
\begin{array}{c}
E[\mathbf{M(r)}]=\frac{\mu_{0}\lambda^{2}}{2}\int_{\Omega}d^{3}r|\nabla\mathbf{M}|^{2}\\
\\\vphantom{}+\frac{\mu_{0}}{2}\int_{\mathbf{R^{\mathrm{3}}}}d^{3}r|\nabla U|^{2}-\mu_{0}\int_{\Omega}d^{3}r\mathbf{H_{\mathrm{e}}\cdot M}.\end{array}\label{eq:energy}\end{equation}
 The three terms above correspond respectively to the exchange energy,
demagnetization (or magnetostatic) energy, and Zeeman energy, with
the (small) magneto-crystalline anisotropy term neglected. (The last
of these can be easily included, but for the materials and geometries
considered here, it is typically overwhelmed by the much larger shape
anisotropy arising from the demagnetization term.) Here $\Omega$
is the volume of the ring, $\lambda=\sqrt{2A/(\mu_{0}M_{s}^{2})}$
is the exchange length (where $M_{s}$, the magnitude of the magnetization,
is assumed to be the same everywhere, and $A$ is the exchange constant),
$\mathbf{H_{e}}$ is the applied external magnetic field, and $|\nabla\mathbf{M}|^{2}\equiv|\nabla M_{x}|^{2}+|\nabla M_{y}|^{2}+|\nabla M_{z}|^{2}$.
The {}``magnetostatic potential'' $U$, arising from long-range
dipole-dipole interactions within the magnetic material, satisfies
$\nabla^{2}U=\nabla\cdot\mathbf{M}$ (and suitable boundary conditions
in the interfaces between media), which can be derived through Maxwell's
equations. Our simulations involve numerical integration of the above
set of equations.

The extremal states of a quasi-$1D$ ferromagnetic ring (i.e., $\Delta R\ll R$
so that the external magnetic field does not vary significantly with
distance from the center of the ring) in a circumferential magnetic
field have been analytically obtained~\citep{martens1}. The solutions
found there apply in the thin-ring limit: $k=t/R\ll1$ and $(\lambda/R)^{2}\sim(t/R)|\ln(t/R)|$.
Under these conditions the second term on the RHS of~(\ref{eq:energy})
separates into three main terms (and a number of smaller ones): a
term which extracts a large energy cost when the magnetization does
not lie completely within the plane of the annulus; a local surface
term (the shape anisotropy, which favors alignment of the magnetization
with the tangential direction at the inner and outer ring radius);
and a nonlocal bulk contribution. Analysis of these terms finds that
the bulk term is small compared to the surface term and can therefore
be neglected~\citep{martens1}. In Sec.~\ref{sec:medium} we test
these conclusions for more realistic geometries by computing numerically
the \textit{total\/} demagnetization energy and comparing it to the
(analytically computable) local surface (i.e., shape anisotropy) term.

In the $1D$ approximation the total energy reduces to \begin{equation}
\begin{array}{c}
E=2\mu_{0}M_{0}^{2}(\frac{\ell}{2\pi})^{2}\frac{t}{R}\\
\\\times\lambda^{2}\int_{0}^{\pi}[(\frac{2\pi}{\ell}\frac{\partial\phi}{\partial\theta})^{2}+\sin^{2}\phi-2h\cos\phi]d\theta,\end{array}\label{eq:reducedenergy}\end{equation}
 where the parameters $\ell$ and $h$ are the scaled circumference
and field: \begin{equation}
\begin{array}{c}
\ell=\frac{R}{\lambda}\sqrt{2\pi\left(\frac{t}{\Delta R}\right)\left|\ln\left(\frac{t}{R}\right)\right|}\\
\\h=\frac{H_{e}}{H_{c}}=\frac{H_{e}}{\frac{\mu_{0}M_{s}}{\pi}\left(\frac{t}{\Delta R}\right)\left|\ln\left(\frac{t}{R}\right)\right|},\end{array}\label{eq:1D}\end{equation}
 and $H_{c}$ is defined below. The first term in the integrand is
the exchange energy, the second the shape anisotropy (i.e., the surface
term arising from the demagnetization energy), and the last is the
Zeeman term.

Given an external magnetic field that is circumferential and points
everywhere in the counterclockwise direction, there are two states
that are local minima of the energy: a stable magnetization configuration
(ground state), which is everywhere aligned with the external field,
and a metastable state that is everywhere antiparallel to the field
(i.e., circumferential and pointing everywhere in the clockwise direction).
$H_{c}$ corresponds in eq.~\ref{eq:1D} to the magnetic field at
which the metastable configuration becomes unstable.

There are also two relevant unstable stationary configurations (i.e.
saddle states). These are defined by the angle $\phi$ that the magnetization
direction makes with the circumferential direction at each point in
the annulus; i.e., $\phi$ is a function of the angle $\theta$ (as
shown in Fig.~\ref{fig:geometry}), $\phi_{\ell,h}(\theta)$. In
the limit of low noise, reversal of the magnetization occurs through
the lower energy saddle state. One of these corresponds to a global
rotation of the magnetization in which $\phi$ is independent of $\theta$;
we therefore label it the {}``constant saddle'', and is denoted~$\phi_{h}$.
The constant saddle favors the exchange and Zeeman energies at the
expense of the demagnetization energy. The second saddle state is
a localized fluctuation of the magnetization and we therefore refer
to it as the {}``instanton saddle'', and denote it by~$\phi_{h,\ell}(\theta)$(denoted
as instanton saddle in \citep{martens1}). This state favors the demagnetization
energy at the expense of the exchange and Zeeman energies.

Which of these two saddles is energetically favored depends on the
applied field and the ring size. When the scaled field $h$ is smaller
than $\sqrt{1-(2\pi/\ell)^{2}}$ the instanton saddle has a lower
energy than the constant saddle; otherwise, the constant saddle is
lower in energy. Fig.~2 of~\citep{martens1} shows the phase boundary
between the two activation regimes as a function of $h$ and $\ell$.

We simulated the dynamics using the analytical solutions as our initial
configurations for the two saddle states: \begin{equation}
\phi_{h}=\cos^{-1}(-h)\label{eq:constant}\end{equation}
 for the constant saddle, and \begin{equation}
\phi_{h,\ell}=2\cot^{-1}\left(\vartheta\ \mathrm{dn}\left(\frac{\theta\mathbf{K}(m)}{\pi}|m\right)\right)\label{eq:soliton}\end{equation}
 for the instanton saddle. Here $\mathrm{dn}(\cdot|m)$ is the Jacobi
elliptic function with $0\leq m\leq1$, and $\mathbf{K}(m)$ is the
complete elliptic integral of the first kind\citep{abramowitzstegun}.
The parameter $m$ satisfies, \begin{eqnarray}
\frac{\ell}{2\mathbf{K}(m)}=\frac{m}{\sqrt{2-m-\sqrt{m^{2}h^{2}+4(1-m)}}}\label{eq:delta}\end{eqnarray}

and $\vartheta$ is defined by,

\begin{equation}
\vartheta=\sqrt{\frac{2-mh-\sqrt{m^{2}h^{2}+4(1-m)}}{2m-2+mh+\sqrt{m^{2}h^{2}+4(1-m)}}}.\end{equation}

For sufficiently small rings $(\ell\leq2\pi)$, the instanton saddle
does not exist (essentially, the variation of the magnetization, which
is of order the exchange length, cannot {}``fit'' onto the ring).
In this limit $m\rightarrow0$ and the instanton solution reduces
to the constant saddle. As the ring becomes larger the parameter $m$
increases from $0$ to $1$ monotonically with $\ell$. For $\ell\gg2\pi$
the $m$ becomes numerically indistinguishable from 1. In this limit
the instanton saddle configuration is given by:

\begin{equation}
\phi_{h,\ell}=2\tan^{-1}\left(\sqrt{\frac{h}{1-h}}\cosh\left(\frac{\theta\ell}{2\pi}\sqrt{1-h}\right)\right).\end{equation}

\begin{figure}
\includegraphics[width=3in]{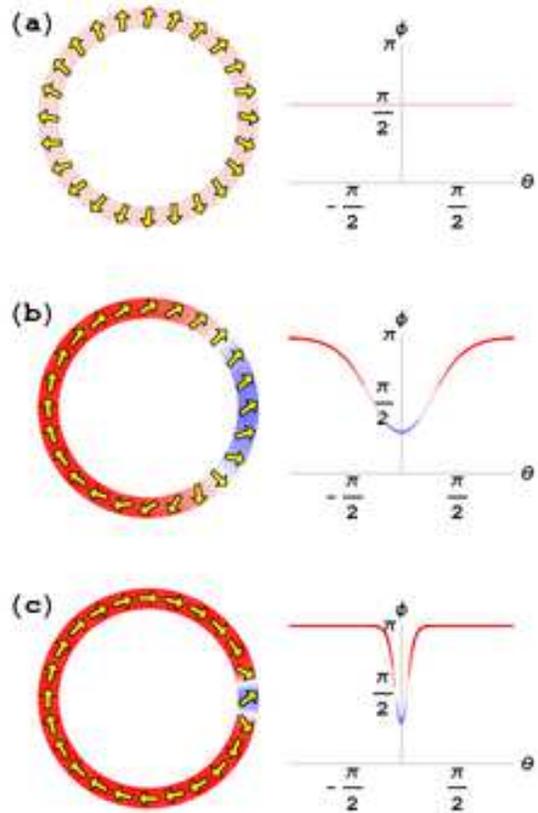}

\caption{\label{fig:configurations}Saddle configurations computed analytically
from the $1D$ model of Martens~\textit{et al.\/} for small, medium
and large ring sizes. a) For small rings the magnetization reversal
goes through the constant saddle state; b) when $\ell>2\pi$ the transition
is via the instanton saddle; c) as the relative size of the ring increases
the fluctuation in the instanton saddle occupies a smaller fraction
of the ring.}

\label{saddleconfigs} 
\end{figure}

We classify rings according to their $\ell$ values as small $(\ell\leq2\pi)$,
medium $(\ell\gtrapprox2\pi)$, and large $(\ell\gg2\pi)$. Physically,
$\ell$ characterizes the ratio of the magnetostatic to exchange energies.
A medium ring $(\ell\gtrapprox2\pi)$ has a scaled circumference close
in size to a domain wall in the material. The saddle configurations
of each regime are shown in Fig.~\ref{saddleconfigs}. %

The activation energies within the 1D analytical model can be calculated
using \ref{eq:reducedenergy}(cf. \citep{martens1}). For the constant
saddle they can be analytically computed: $\Delta E=\mu_{0}M_{0}^{2}t^{2}R|\ln(t/R)|(1-h)^{2}$.

\section{Method}

\label{method}

We studied thin nanorings by running simulations on the model of Sect.~\ref{sec:model}
using the publicly available packages OOMFF and Nmag~\citep{donahue1999,Fischbacher2007}.
These packages effectively simulate the dynamics specified by~(\ref{eq:LLG})
and (\ref{eq:energy}) at zero temperature; i.e., all runs start from
an initial configuration and run downhill in energy.

Our initial states were the instanton and constant saddles described
by~(\ref{eq:constant}) and (\ref{eq:soliton}), which provided starting
points that were guessed to be relatively close to the actual saddles.
The system subsequently relaxed to the actual saddle states, which
turned out to be remarkably close to analytical solutions. We describe
below how this was determined.

Depending on the starting state, the system will evolve to one or
the other (meta)stable state, i.e., either the clockwise or the counterclockwise
magnetization configuration. In order to find the actual $2D$ saddle
numerically (recall that the magnetization is forced by the magnetostatic
energy term to lie in the plane of the ring), we introduce a new field
value, denoted $h_{t}$, an {}``effective field'' for which the
state $\phi_{h,l}$ behaves as a saddle state. In determining $h_{t}$,
two criteria are used. First, $\phi_{h,l}$ must be as nearly a stationary
state as the numerics allow, i.e., the initial time derivative of
the total micromagnetic energy should be close to zero $(\lim_{h_{e}\rightarrow h_{t}}\left.\frac{\delta E}{\delta t}\right|_{t=0}\
\rightarrow0^{-})$. Second, the state $\phi_{h,l}$ should mark the boundary between
the basins of attraction of each (meta) stable state (i.e., for $h<h_{t}$
the system evolves to a clockclockwise state, while for $h>h_{t}$,
it evolves to the counterclockwise state). In the following sections,
we show that these criteria are satisfied in a variety of rings with
different exchange lengths and annular widths. We also show how the
model eventually breaks down when the width of the ring becomes very
large.

The procedure can be summarized is as follows. For a given initial
$\phi_{h,l}$ we find the appropriate $h_{t}$ by a bracket and bisection
iterative process. We set the initial configuration $\phi_{h,l}$,
fix the external magnetic field at the value $h_{e}$ and allow the
system to relax. If the final state is the metastable (clockwise)
configuration we increase $h_{e}$ by an amount $\delta h_{e}$; if
the final state is the stable (counterclockwise) configuration we
decrease $h_{e}$ by $\delta h_{e}$. We then start a new run and
reduce $\delta h_{e}$ by a factor of $2$. As $\delta h_{e}$ decreases,
the total relaxation time increases due to the slow dynamics at the
start of the simulation, providing evidence that the initial configurations
are approaching the true saddle states. We iterate until we reach
a numerical uncertainty of $\delta h=6\times10^{-3}$.

The Nmag simulations were run using a mesh consisting of 20963 volume
elements, 15154 surface elements, and 7594 points with an edge length
of average 3.89 nm and standard deviation of 0.7. (The quality distribution
of the mesh was 1.66\% below 0.6; 9.11\% between 0.6 and 0.7, 57.28\%
between 0.7 and 0.8, 32.05\% between 0.8 and 0.9, and 0.01\% above
0.9.) The cell sizes for the simulations in OOMMF were selected so
that they were smaller than the exchange length for each of the regimes
studied. Different values of $\ell$ for a given geometry were studied
by changing the exchange constant $A$ (cf.~below~(\ref{eq:energy}))
and keeping the ring dimensions constant. This changes $\lambda$
and therefore $\ell$, the mesh can be adjusted to speed up the simulations.

\subsection{String Method in Rare Events}

The String Method \citep{Weinan2002} is a recently introduced numerical
procedure for calculating transition energies and paths within the
context of large fluctuations and rare events. It is useful to find
the path connecting two stable configurations $\mathbf{M_{A}}$ and
$\mathbf{M_{B}}$, through a curve $\xi$ with minimum energy. The
obtained path corresponds to the reversal trajectory in the limit
where the precession term of~(\ref{eq:LLG}) is negligible compared
to the damping term. This curve $\xi$ satisfies

\begin{equation}
\nabla_{\mathbf{M}}E^{\perp}(\xi)=\nabla_{\mathbf{\mathbf{M}}}E(\xi)-[\nabla_{\mathbf{\mathbf{M}}}E(\xi)\cdot\hat{t}]\hat{t}=0\end{equation}
 where $\hat{t}$ is the unit tangent of the curve $\xi$. The curve
$\xi$ is found by guessing a parametrized path $\xi(0)=\{\mathbf{M}(\alpha)\in[0,1],\mathbf{M}(0)=\mathbf{M_{A}},\mathbf{M}(1)=\mathbf{M_{B}}\}$
and evolving it in {}``time'' according to:

\begin{equation}
\partial_{t}\mathbf{M}(\alpha)=-\nabla_{\mathbf{M}}E^{\perp}(\mathbf{M}(\alpha))+\lambda\hat{t.}\end{equation}
 The second term is added to enforce a particular parametrization;
it does not alter the actual evolution of the curve. It is convenient
to rewrite this equation as:\begin{equation}
\partial_{t}\mathbf{M}(\alpha)=H_{\mathrm{eff}}^{\perp}(\mathbf{M}(\alpha))+\lambda\hat{t}\label{eq:stringdifeq}\end{equation}
 For numerical purposes the path $\xi$ is discretized with $N+1$
points betwen $\mathbf{M_{A}}$and $\mathbf{M_{B}}$. After each iteration
of (\ref{eq:stringdifeq}) with an Euler forward algorithm the magnetization
vectors are renormalized to $\mathbf{M_{s}}$. In Sect. \ref{sec:wide}
we use this method to find the barrier between two states connected
through a transition state.

\section{Medium Size Narrow Ring}

\label{sec:medium}

\begin{figure}
\includegraphics[width=3in]{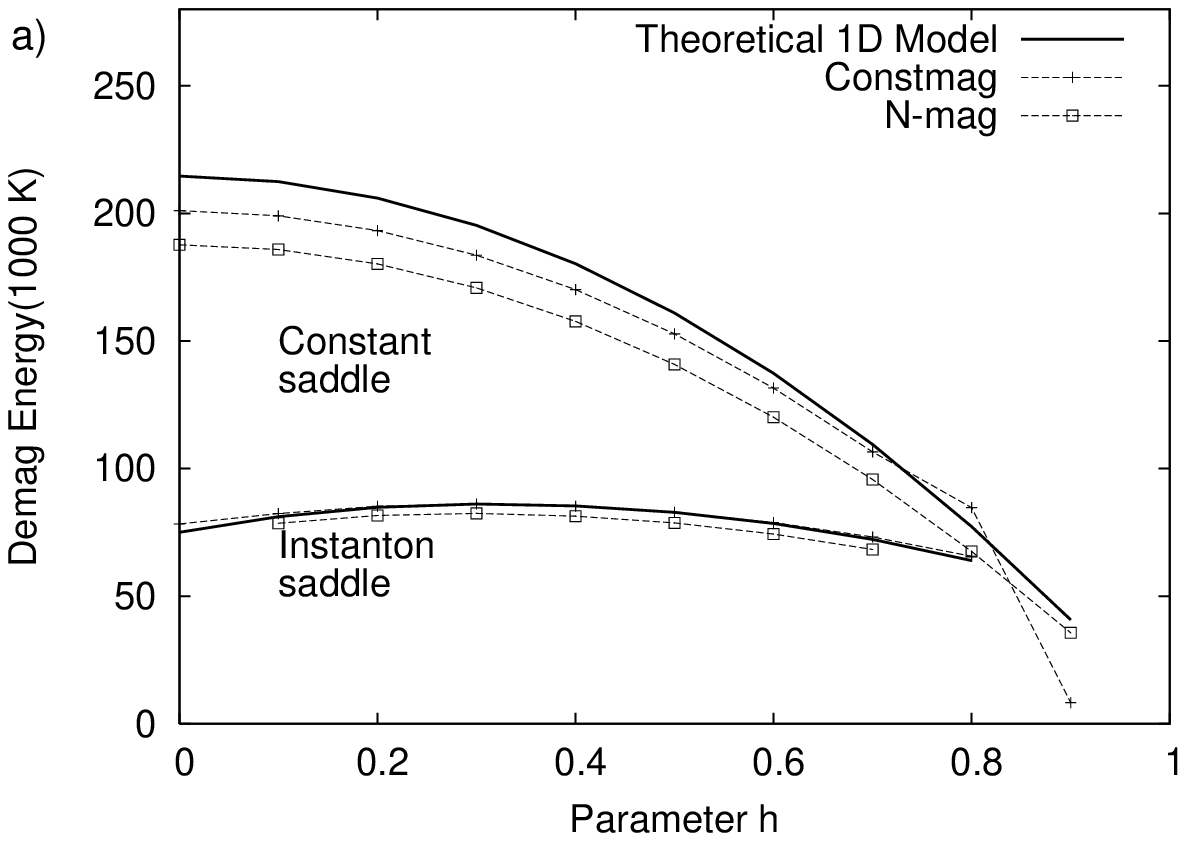}

\includegraphics[width=3in]{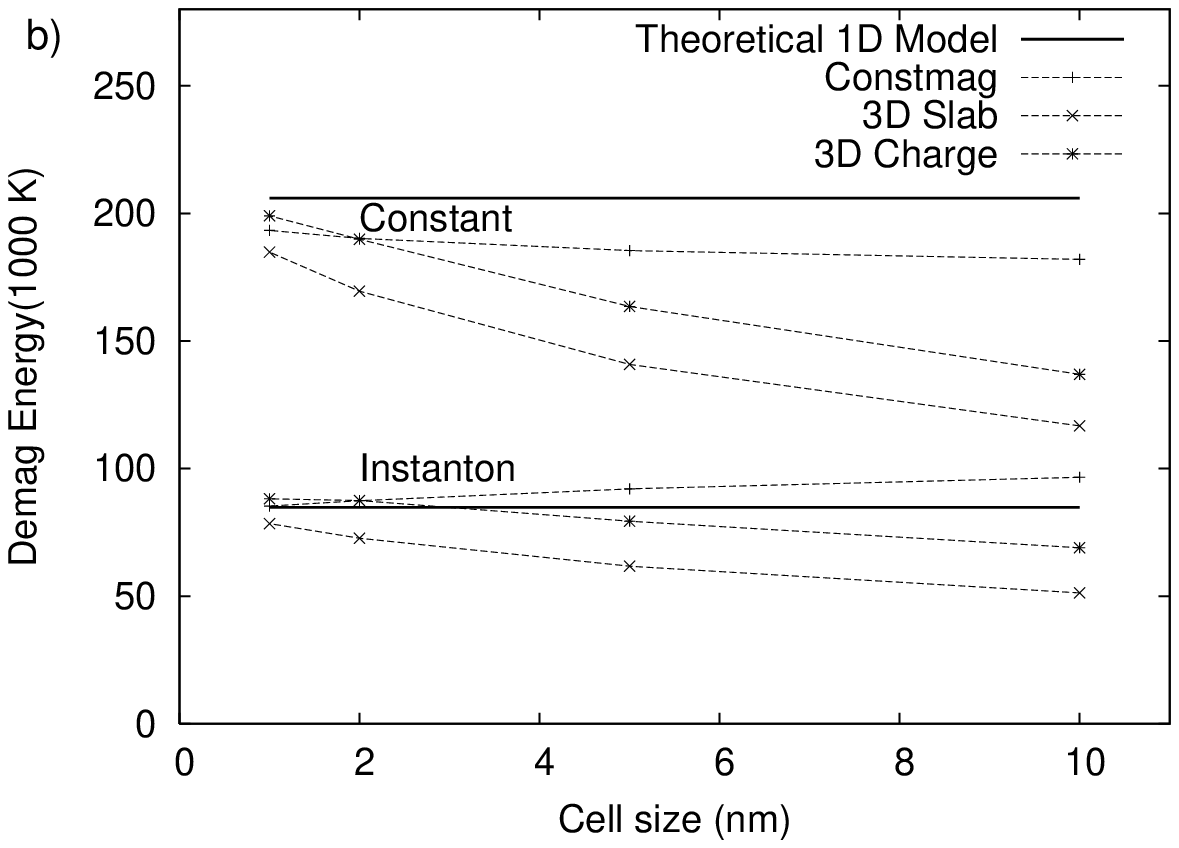}

\caption{Total demagnetization energies, using different calculation methods,
compared to analytical results for the shape anistropy term alone
($A=3.2\times10^{-10}$ J/m, $R=200$ nm , $t=2$ nm and $\Delta R=40$
nm). a) Comparison of OOMMF using the ConstMag method at cell size
$1$ nm and Nmag with theoretical predictions vs. $h$. b) Demagnetization
energy vs. cell size using three different methods for computing this
energy in OOMMF, all for $h=0.2$. The solid line is the analytical
computation of the shape anisotropy term alone.}

\label{fig:demagenergies} 
\end{figure}

We consider a narrow ring of medium reduced circumference $\ell$
(i.e., the parameter $m$ not close to 1) with $\lambda/R\ll1,$ $t/R\ll1$
and $(\lambda/R)^{2}\sim(t/R)|\ln(t/R)|$ ($A=3.2\times10^{-10}$
J/m, $\Delta R=40$ nm, $R=200$ nm, $t=2$ nm, $M_{s}=8\times10^{5}$
A/m). With these values, $\ell$ and $H_{c}$ are $12$ and $73.9$
mT, respectively. We first test numerically whether the surface, or
shape anisotropy, term is in fact the main contributor to the magnetostatic
energy, as required for the vailidity of the analytic solutions to
hold~\citep{martens1}. Using OOMMF and Nmag the total demagnetization
energy were obtained for different values of $h$ and compared to
the values of the surface term (second term in the integrand of~(\ref{eq:reducedenergy})).
The results of this comparison are presented in Fig.~\ref{fig:demagenergies},
which %

shows that the numerical computation of the total demagnetization
energy gives nearly the same dependence on $h$ as the contribution
from the shape anisotropy alone. This provides numerical support for
the approximations used to arrive at the $1D$ analytical solutions
of~\citep{martens1} and confirms that the bulk magnetostatic term,
neglected in the analytic model, is indeed not important.

We also compare the total demagnetization energies computed using
OOMMF for different cell sizes in Fig.~\ref{fig:demagenergies}b.
Three suitable methods to calculate the demag energy are available
in the OOMMF package; ConstMag, 3dSlab, 3dCharge. Constmag calculates
the average demagnetization field assuming the magnetization is constant
in each cell; 3dSlab uses a demag field obtained from blocks of constant
charge; and 3dCharge uses constant magnetization to calculate the
in-plane component of the magnetic field, and constant charges to
calculate the out of plane demagnetizing field. As seen in the figure,
the consistency between different methods of calculation improves
as cell size is reduced; and the numerical results approach that of
the $1D$ model.

\begin{figure}
\includegraphics[width=3in]{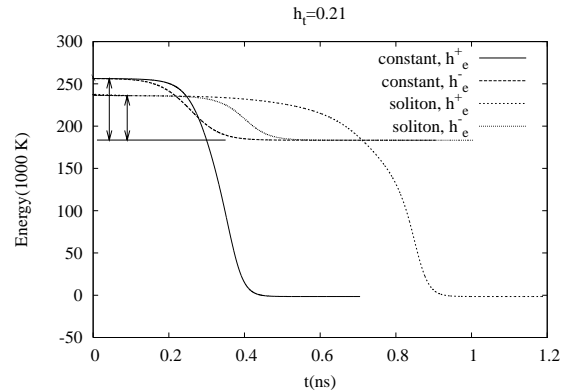}

\caption{\label{fig:dynamicssample}Evolution of the total micromagnetic energy
with time. Arrows show activation energy of each saddle. The configurations
shown bifurcates at the fields given and have an initial slow evolution.
The two criteria used to determine a saddle state.}

\end{figure}

Once $h_{t}$ is obtained following the method described in Sect.~\ref{method},
the saddle state is numerically obtained and the activation energy
is thereby determined from the difference between each saddle state
and the metastable state. Fig.~\ref{fig:dynamicssample} displays
curves at $h_{t}=0.21$ for each of the two saddle states used as
initial configurations.%

As can be seen in the figure, after a very short transient the system
arrives at a configuration in which the energy stays almost constant
for an extended period; this indicates that the initial $1D$ analytical
solution is close in both energy and its geometrical configuration
to the true $2D$ saddle. Eventually, the saddle state decays into
one of the two stable configurations. The activation energy is easily
computed this way (as seen in Fig.~\ref{fig:dynamicssample}), and
a glance at the figure confirms that for the applied field $h_{t}=0.21$
the instanton configuration has a lower activation energy than the
constant saddle, as predicted theoretically~\citep{martens1}.

\begin{figure}
\includegraphics[width=3in]{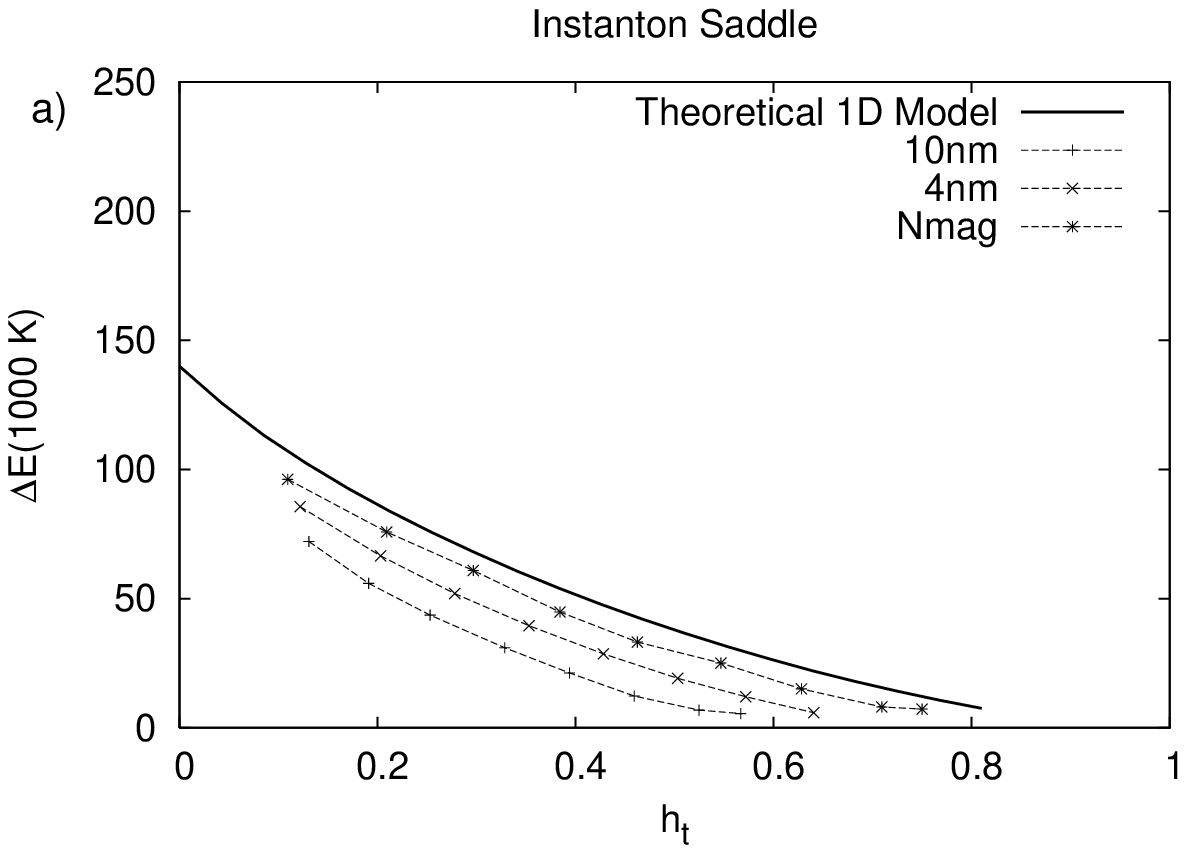}\smallskip{}
 \includegraphics[width=3in]{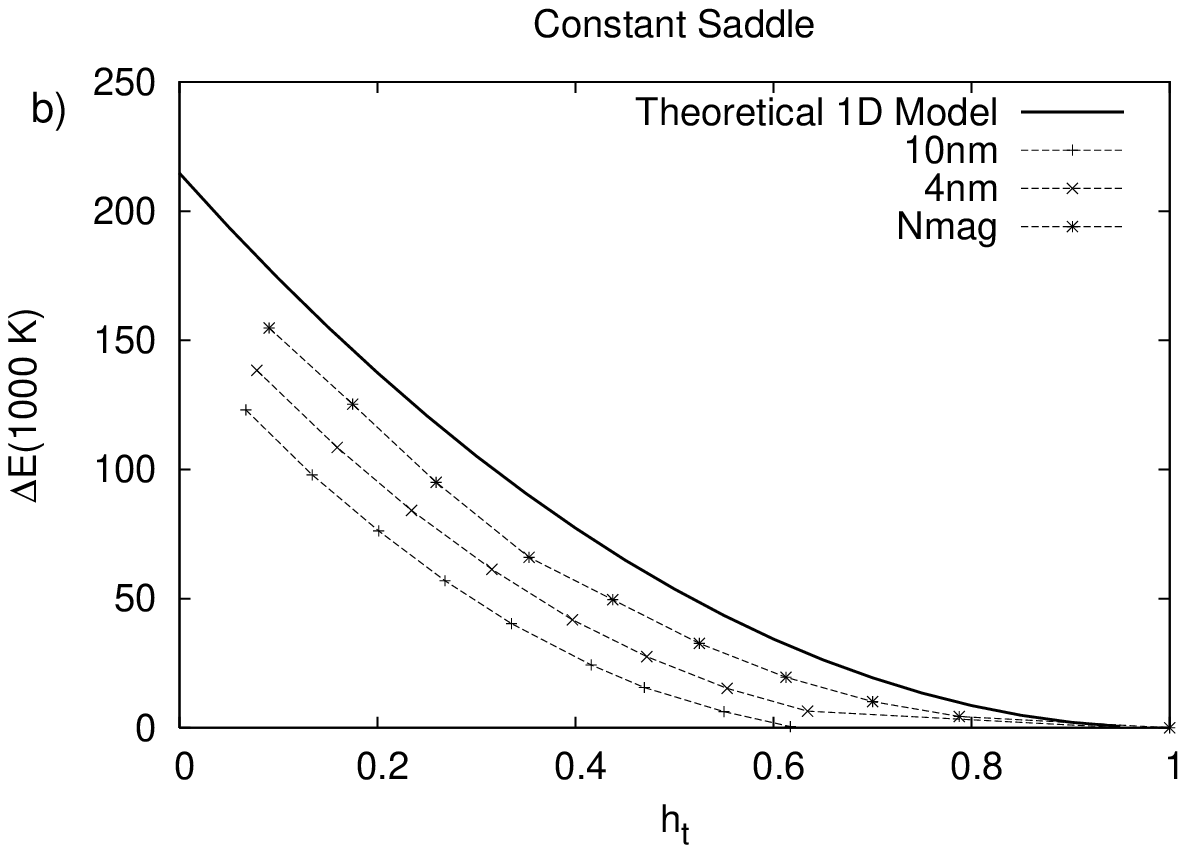}\smallskip{}
 \includegraphics[width=3in]{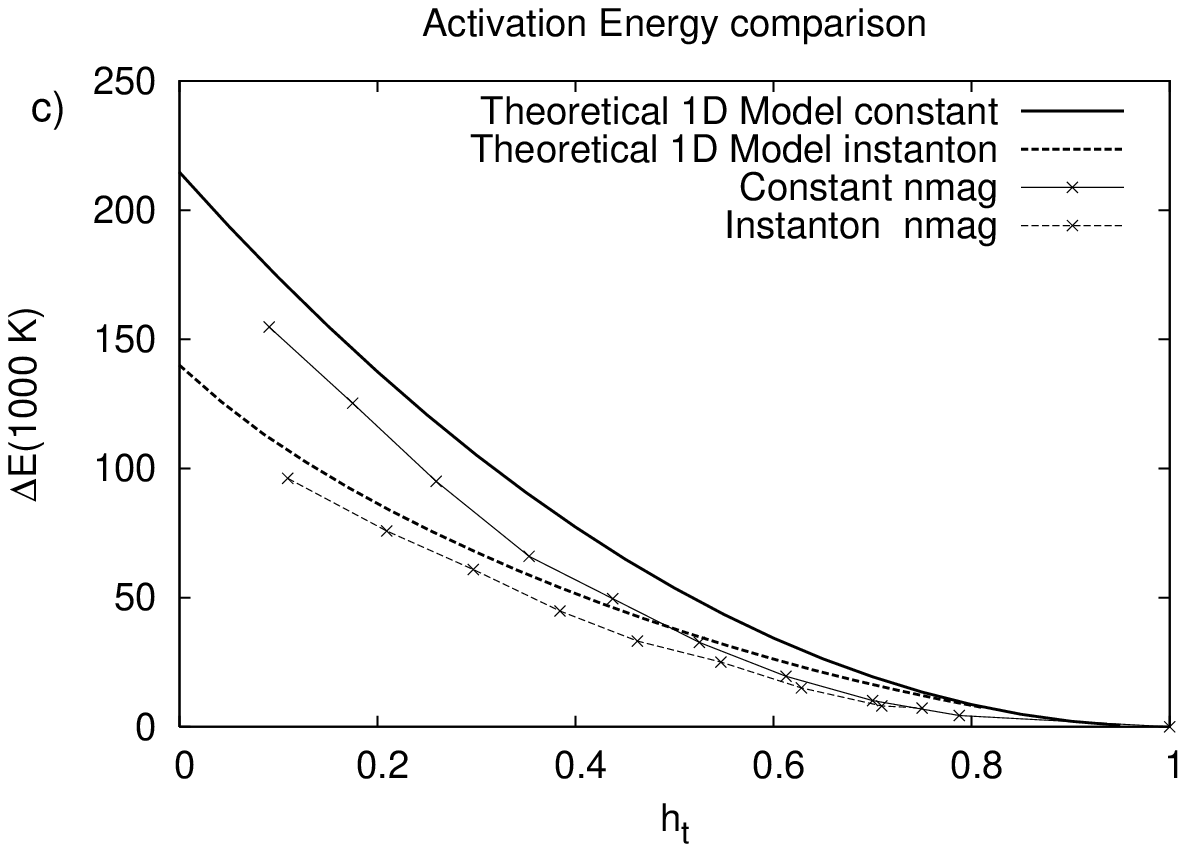}

\caption{\label{fig:midnarrowring}Comparison of the activation energies for
the (a) instanton and (b) constant saddles. Data points represent
steps of $0.1$ in $h$. As the numerics are refined both by using
Nmag and by reducing the cell size in OOMMF simulations the results
approach the theoretical predictions. (c) Comparison for Nmag for
the middle size, narrow ring ($\ell$=12, $\Delta R$=40 nm) of the
corresponding curves shown in (a) and (b).}

\end{figure}

The method described in Sect.~\ref{method} was repeated to obtain
the behavior of $\Delta E$ as a function of $h_{t}$. Using this
approach one can calculate $h_{t}\approx\frac{h_{e}^{+}+h_{e}^{-}}{2}$and
$\Delta E=\frac{E[\phi_{h},h_{e}^{+}]+E[\phi_{h},h_{e}^{-}]}{2}-E[\phi=\pi,h_{t}]$
for each of the two saddle configurations. Here, $E[\phi_{h},h_{e}]$
represents the numerical energy of a configuration $\phi_{h}$ under
an applied field of magnitude $h_{e}$. The results are summarized
in Fig.~\ref{fig:midnarrowring} together with the analytical predictions.

Figs.~\ref{fig:midnarrowring}a and \ref{fig:midnarrowring}b show
$\Delta E(h_{t})$ for the instanton saddle and constant saddle, respectively.
From bottom to top the curves represent calculations in OOMMF for
two different cell sizes, an Nmag calculation and the analytical prediction.

The Nmag results are closer to the analytical predictions than the
OOMMF calculations. This is due to the fact that the curvature is
more faithfully represented by a mesh of tethrahedrons in Nmag, whereas
OOMMF represents the ring with a square grid. The edges of the tethrahedrons
can be arranged to follow closely the curvature of the ring, whereas
the square cells edges will in most cases make a finite angle with
the ring tangent. This results in a much larger error contribution
to the demagnetization energy in square grids in OOMMF than in Nmag.

Fig.~\ref{fig:midnarrowring}c presents the Nmag simulation results
for the activation energy. As predicted in Ref.~\citep{martens1},
the instanton saddle configuration is preferred at lower fields. For
a fixed $\ell$ the activation energy curves are predicted to cross
at $h_{c}(\ell)=\sqrt{1-(2\pi/\ell)^{2}}$. For higher fields, the
constant configuration is the sole saddle state (this is in contrast
to the low-field side, where both solutions exist but the constant
saddle has higher energy). Numerically, the field at which the crossover
between the saddles occurs is somewhat lower than that predicted.
This discrepancy arises because the theoretical model applies to a
strictly $1D$ ring, whereas the simulations run using higher-dimensional
rings (2D in OOMMF and 3D in Nmag).

\section{large size narrow ring}

\begin{figure}
\includegraphics[width=3in]{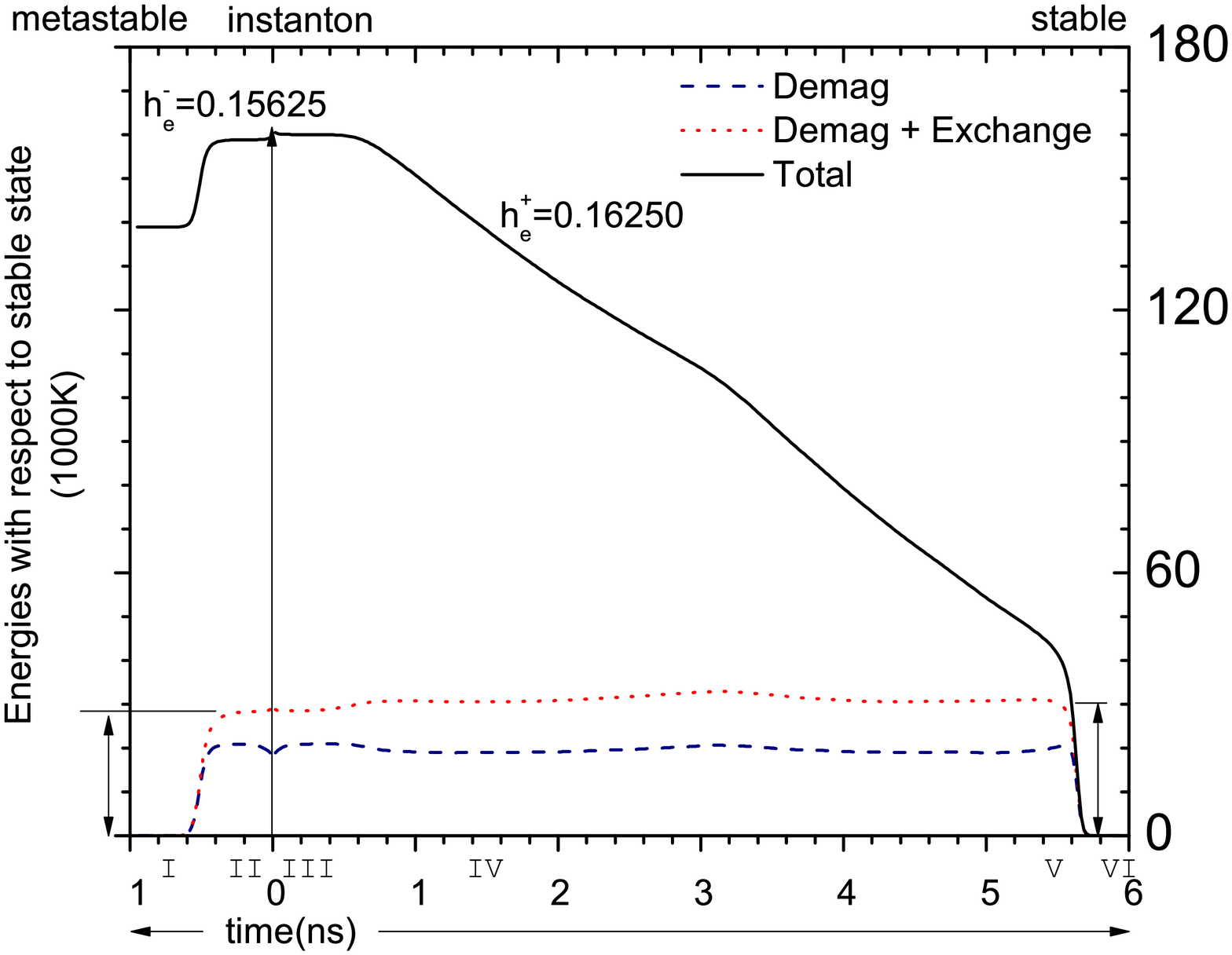}\medskip{}

\includegraphics[width=1in]{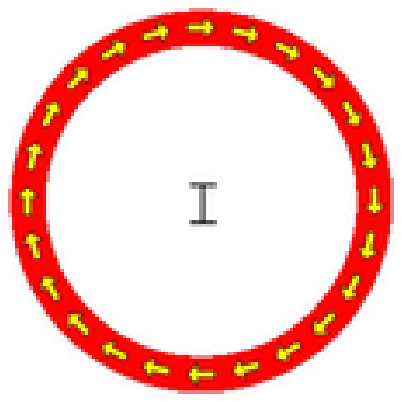}\includegraphics[width=1in]{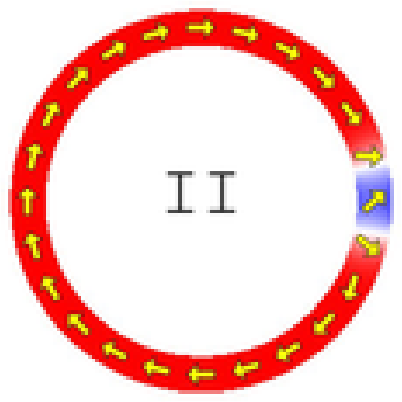}\includegraphics[width=1in]{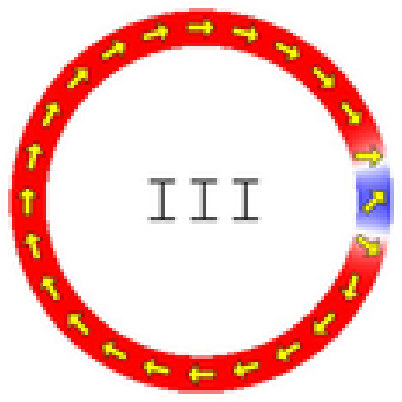}\medskip{}
 \includegraphics[width=1in]{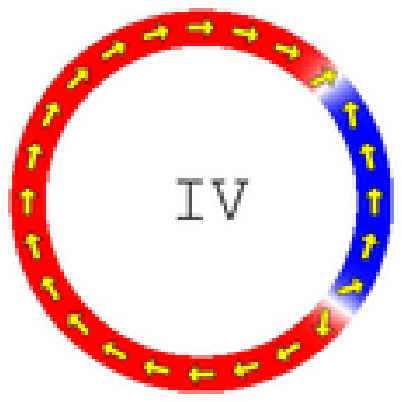}\includegraphics[width=1in]{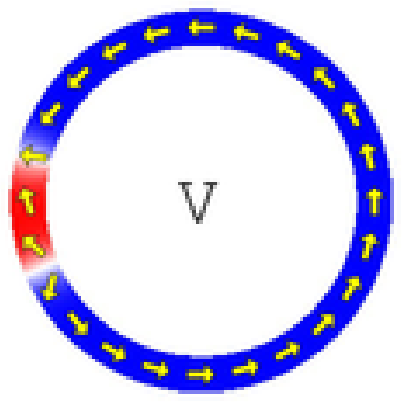}\includegraphics[width=1in]{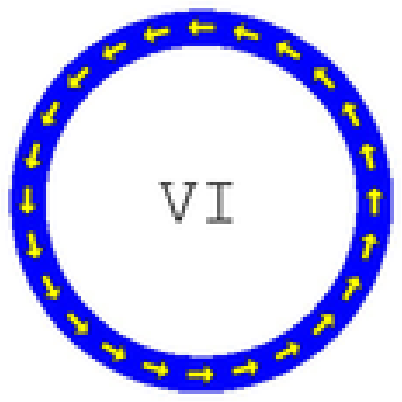}

\caption{\label{fig:largeringsoliton}Evolution of the total energy with time
for the instanton saddle, with energies measured with respect to that
of the stable state. The decay to the metastable state is shown as
proceeding to the left of $t=0$, and to the stable state to the right
of $t=0$. The ring dimensions are $\Delta R=40$ nm, $\ell=60$.
When $h_{e}<h_{t}$ the instanton saddle (II close to $t$=0) decays
quickly into the metastable configuration (I at $t$=0.7 ns); domain
walls are annihilated. When $h_{e}>h_{t}$ the two domain walls of
the instanton saddle (III. close to $t$=0) separate (IV, t=1.5ns),
move to the opposite side of the ring (V, $t$=5.4ns) and annihilate
(VI, $t$=5.7ns).}

\end{figure}

\begin{figure}
\includegraphics[width=3in]{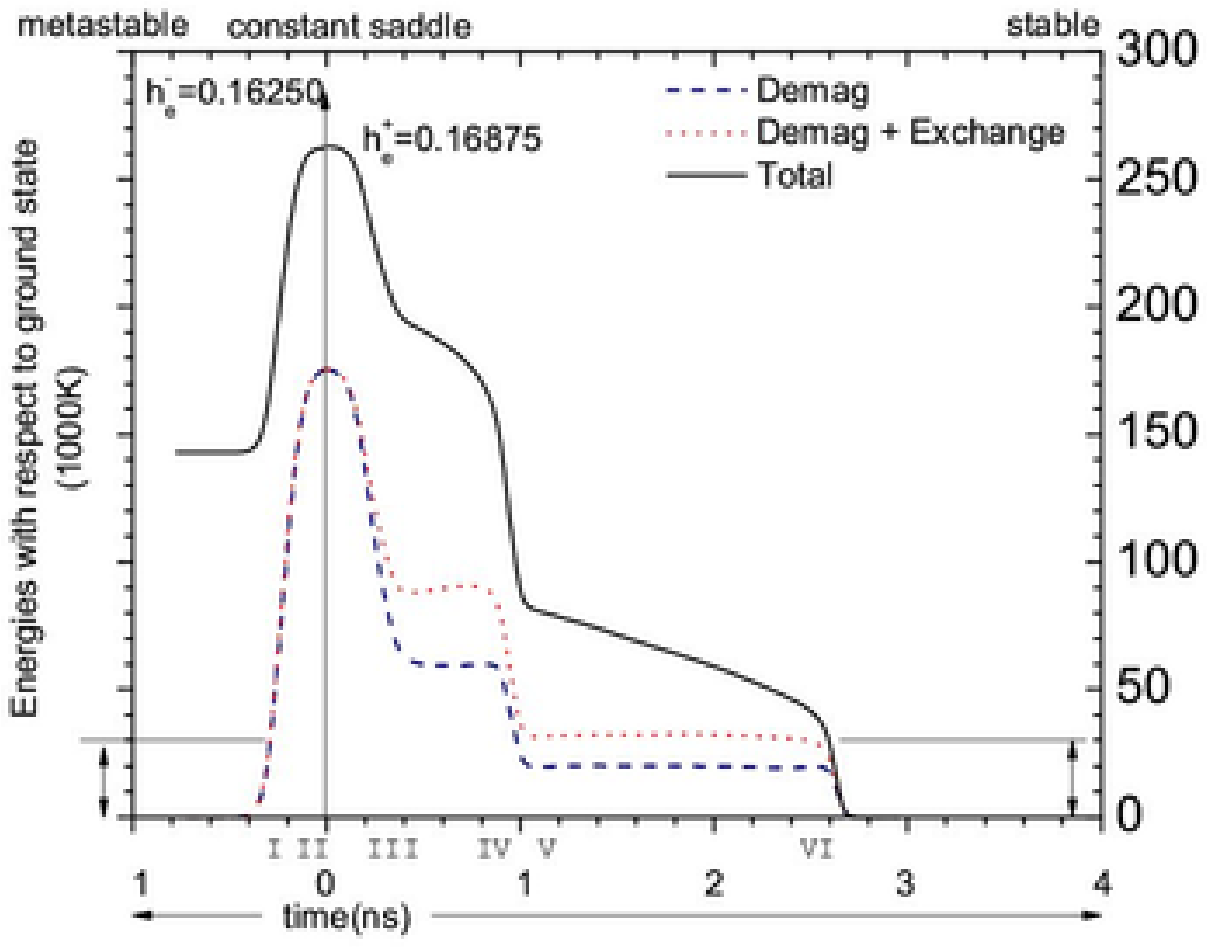}\medskip{}
 \includegraphics[width=1in]{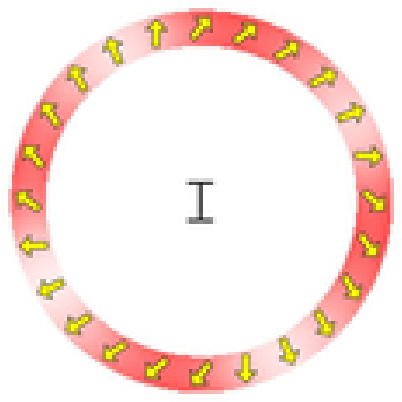}\includegraphics[width=1in]{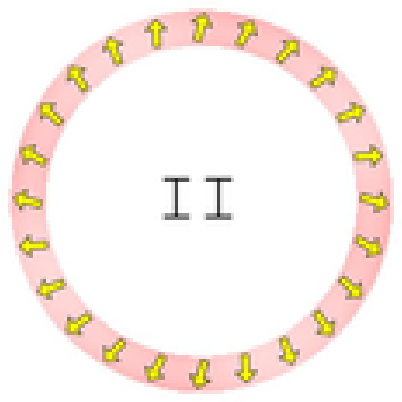}\includegraphics[width=1in]{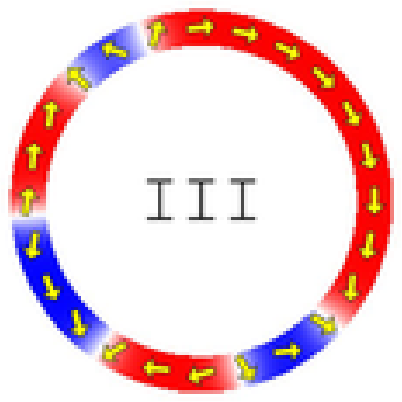}\medskip{}
 \includegraphics[width=1in]{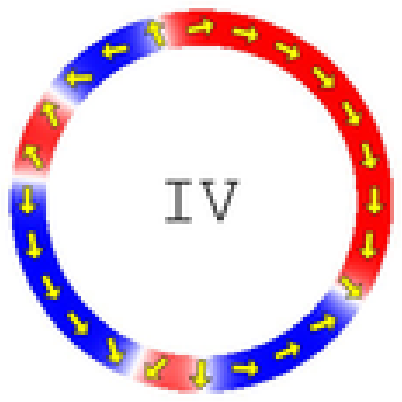}\includegraphics[width=1in]{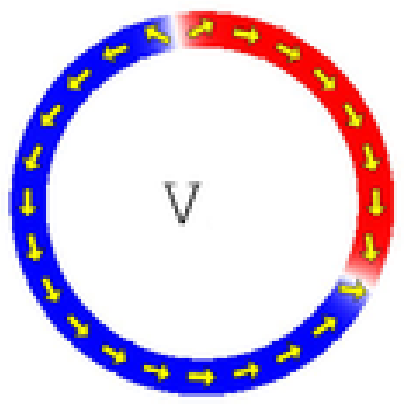}\includegraphics[width=1in]{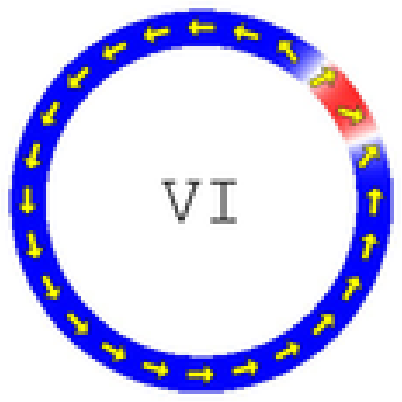}

\caption{\label{fig:constantm1dynamics}Evolution of the total energy with
time for the constant saddle, with energies measured with respect
to that of the stable state. The decay to the metastable state is
shown as proceeding to the left of $t=0$, and to the stable state
to the right of $t=0$.The ring dimensions are $\Delta R=40$ nm,$\ell=60$.
When $h_{e}<h_{t}$ the constant saddle (II at $t$=0.15ns) starts
to decay into the metastable configuration (I at $t$=0.28 ns), it
is possible to see that this decay is not uniform in all the ring,
and some regions decay faster than others. When $h_{e}>h_{t}$ the
constant saddle quickly develops domains (III at $t$=0.51ns). Counterclockwise
domains expand at the expense of clockwise domains (IV, $t$=0.83ns),
with annihilitation that produce a sudden decrease in energy (V, $t$=1.03ns)
leaving a single domain wall pair which continues moving (VI, $t$=2.52ns)
until eventual annihilation; the slope during the last stage is very
close in magnitude to the slope of Fig. \ref{fig:largeringsoliton}
during domain wall propagation, the discrepancy is due to the difference
of applied fields.}

\end{figure}

We now investigate an annular film with the same dimensions but different
$\lambda$ (e.g., permalloy with $A=1.3\times10^{-11}J/m$); the rest
of the dimensions are kept the same as above. Such a ring belongs
to the large ring regime $(\ell=60,m\approx1)$. As before, we can
obtain a qualitative understanding of the reversal process by following
the time evolution of the micromagnetic energy; this is presented
in Figs.~\ref{fig:largeringsoliton},\ref{fig:constantm1dynamics}.
Using the fact that an instanton is a superposition of two domain
walls with opposite chirality~(Fig.~\ref{fig:highmetastable}b)\citep{Braun},
the features observed in Figs.~\ref{fig:largeringsoliton}, \ref{fig:constantm1dynamics}
can be explained and an intuitive idea of the reversal mechanism developed.

\begin{figure}
\includegraphics[width=3in]{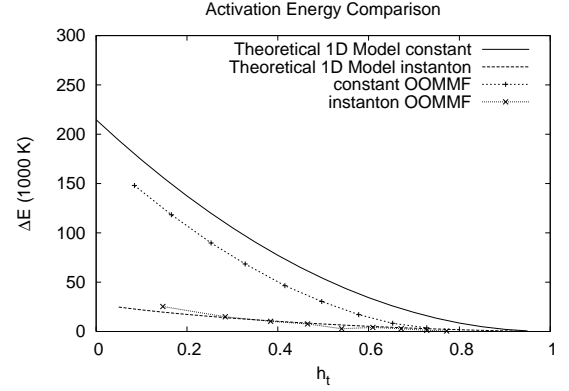}

\caption{\label{fig:m1activationenergies}Total energies of the two saddle
states for dimensions $\Delta R=40$ nm,$\ell=60$ with respect to
the clockwise configuration.}

\end{figure}

We begin by discussing the evolution of energy with time as shown
in Fig.~\ref{fig:largeringsoliton}a. For $h_{e}<h_{t}$ the instanton
decays into the metastable state with a rapid decrease in energy (seen
to the left of $t=0$). This corresponds to the annihilation of two
transverse domains walls. The field in this case is not sufficient
to separate the two transverse walls. When $h_{e}>h_{t}$ (right side)
the energy of the system decays linearly with time towards the stable
state and then sharply decreases. The linear time decay corresponds
to movement of the transverse domain walls in opposite directions
(due to the external field) at approximately constant speed as can
be seen in Fig. \ref{fig:largeringsoliton}. The final, sharp decrease
in energy results from the collision and annihilation of the walls.
Note that the magnetostatic and exchange energies remain almost constant
while the walls are propagating (roughly) independently of each other.
The annihilation energies of the domain walls (indicated by the arrows
in Fig.~\ref{fig:largeringsoliton}a) are roughly the same magnitude.

The slope of the $E(t)$ curve during the propagation phase provides
a measure of how fast the reversal proceeds. For the particular damping
parameter used, this can be used to estimate the wall speed from the
last term of (\ref{eq:reducedenergy}) to be $v=\frac{1}{8n\mu_{0}M_{0}^{2}}(\frac{2\pi}{\ell})^{2}\frac{R^{2}}{\lambda^{2}t}\langle\frac{dE}{dt}\rangle$where
$n$ is the number of domain wall pairs present in the ring. In deriving
this expression $\phi=0,\pi$ for each domain and domain wall widths
are assumed to remain constant during propagation. A comparison to
medium-sized rings shows that at fixed $h$ the reversal time increases
as $\lambda$ decreases for two reasons: the domain wall width is
comparatively smaller and the effective scaled circumference $\ell$
is comparatively larger (cf.~Figs.~\ref{fig:dynamicssample} and
\ref{fig:largeringsoliton}).

The time dependence of the energy for the constant saddle case, Fig.~\ref{fig:constantm1dynamics},
is seen to proceed in several steps where the energy decrease is gradual,
punctuated by large changes in the slope $dE/dt$. These features
can be explained as follows. Fig.~\ref{fig:m1activationenergies}
shows that the activation energy of the constant saddle at $h\approx0.17$
is several times larger than the activation energy of the instanton
saddle at the same field. With such a large activation energy, it
is relatively easy to create several domain wall pairs along different
(randomly placed) parts of the ring, each of them with an energy cost
roughly equal to the activation energy of the instanton. In the simulations,
this process is modified by the discretization of the ring, and in
an experimental setup it is possible that impurities and edge roughness
might play a similar role. The abrupt changes in slope are due, as
before, to the annihilation of domain wall pairs and hint at a richer
spectrum of states that are stationary in the energy when the scaled
circumference is large compared to the exchange length.

\subsection{Multiple Wall Pair Spectra}

\label{sec:multiple}

We have also gone beyond the work of Martens~\textit{et al.\/} by
finding numerically new stable states consisting of multiple domain
wall pairs. When present, these states influence the time evolution
of the system in a manner similar to that just described. In this
section we discuss these new states and present a model to incorporate
their effects on magnetization reversal.%
\begin{figure}
\includegraphics[width=3in]{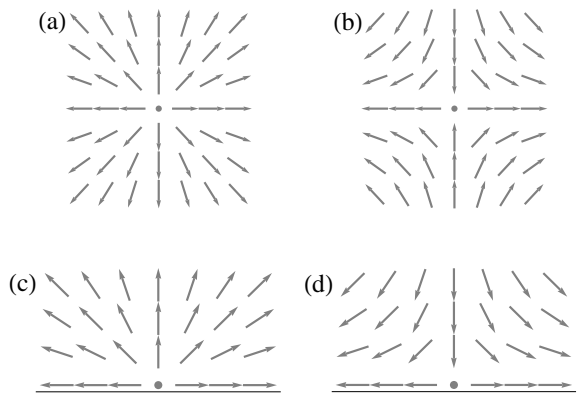}

\caption{\label{fig:singularities}Topological defects of the XY model in the
bulk (a,b) and in the presence of an edge (c,d), with winding numbers
of 1,-1,+1/2,-1/2 respectively.}

\end{figure}

\begin{figure}
\includegraphics[width=1in]{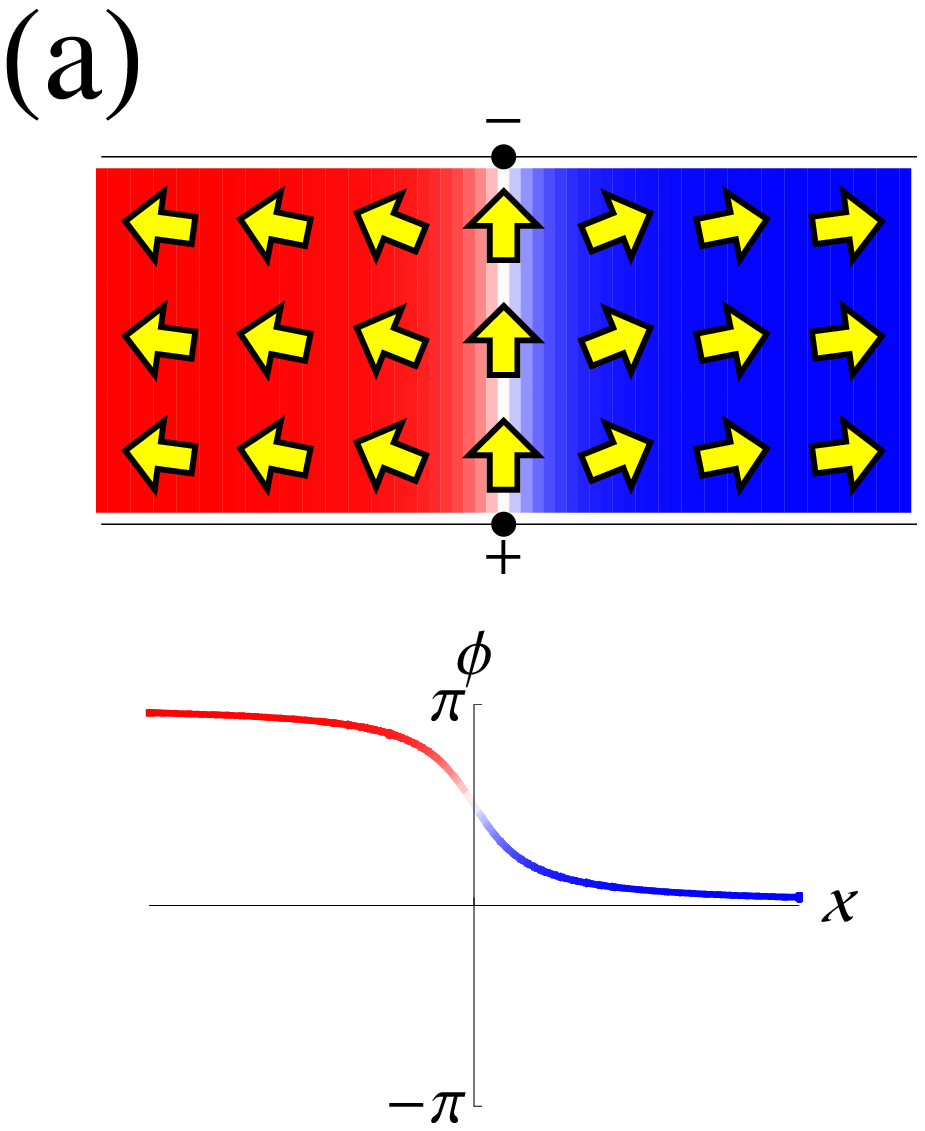}\includegraphics[width=1in]{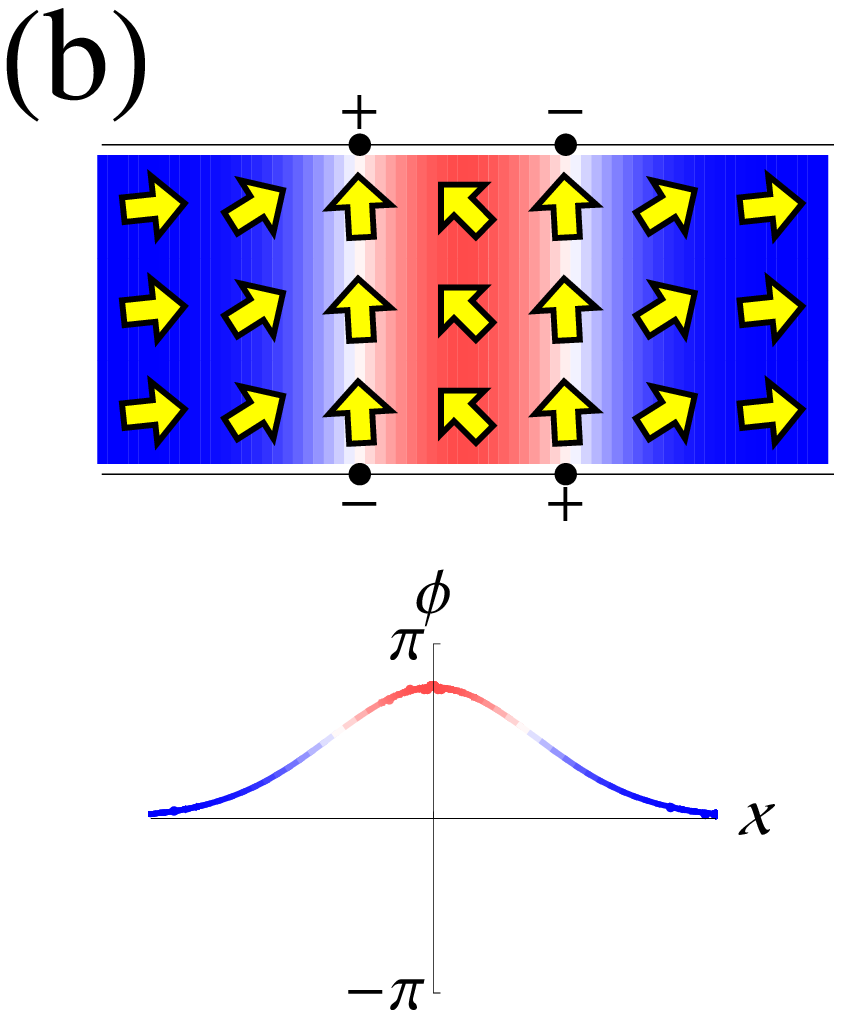}\includegraphics[width=1in]{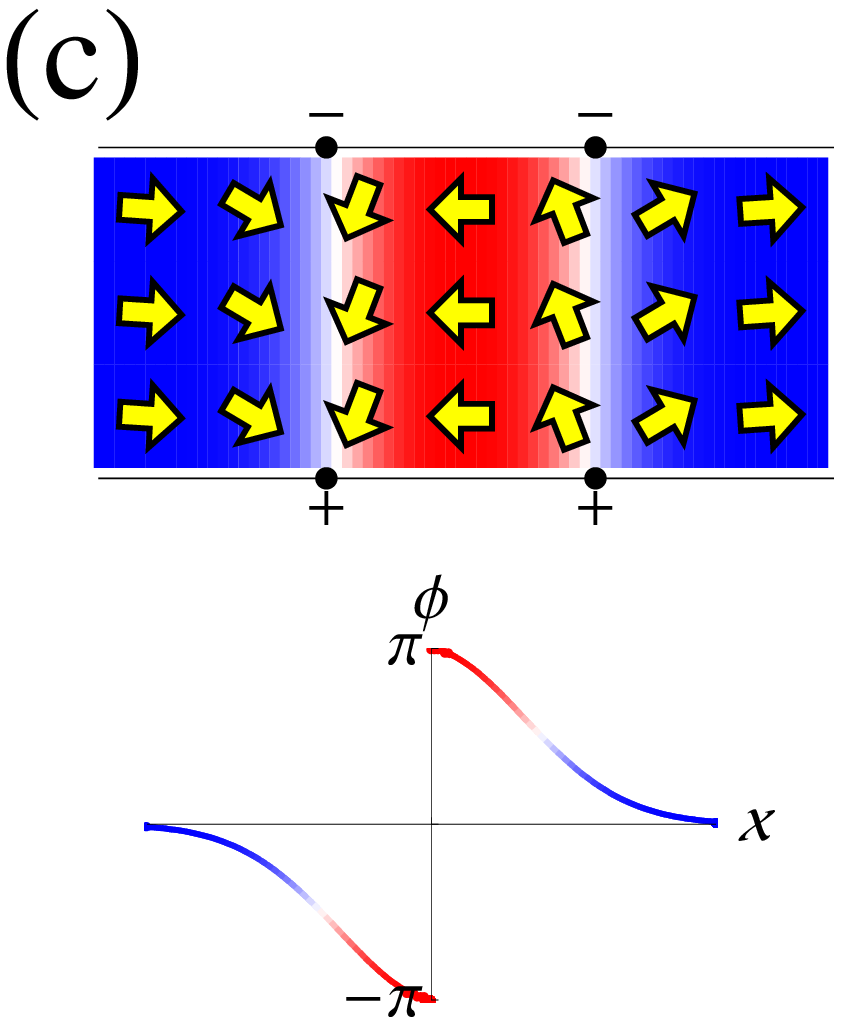}

\includegraphics[width=3in]{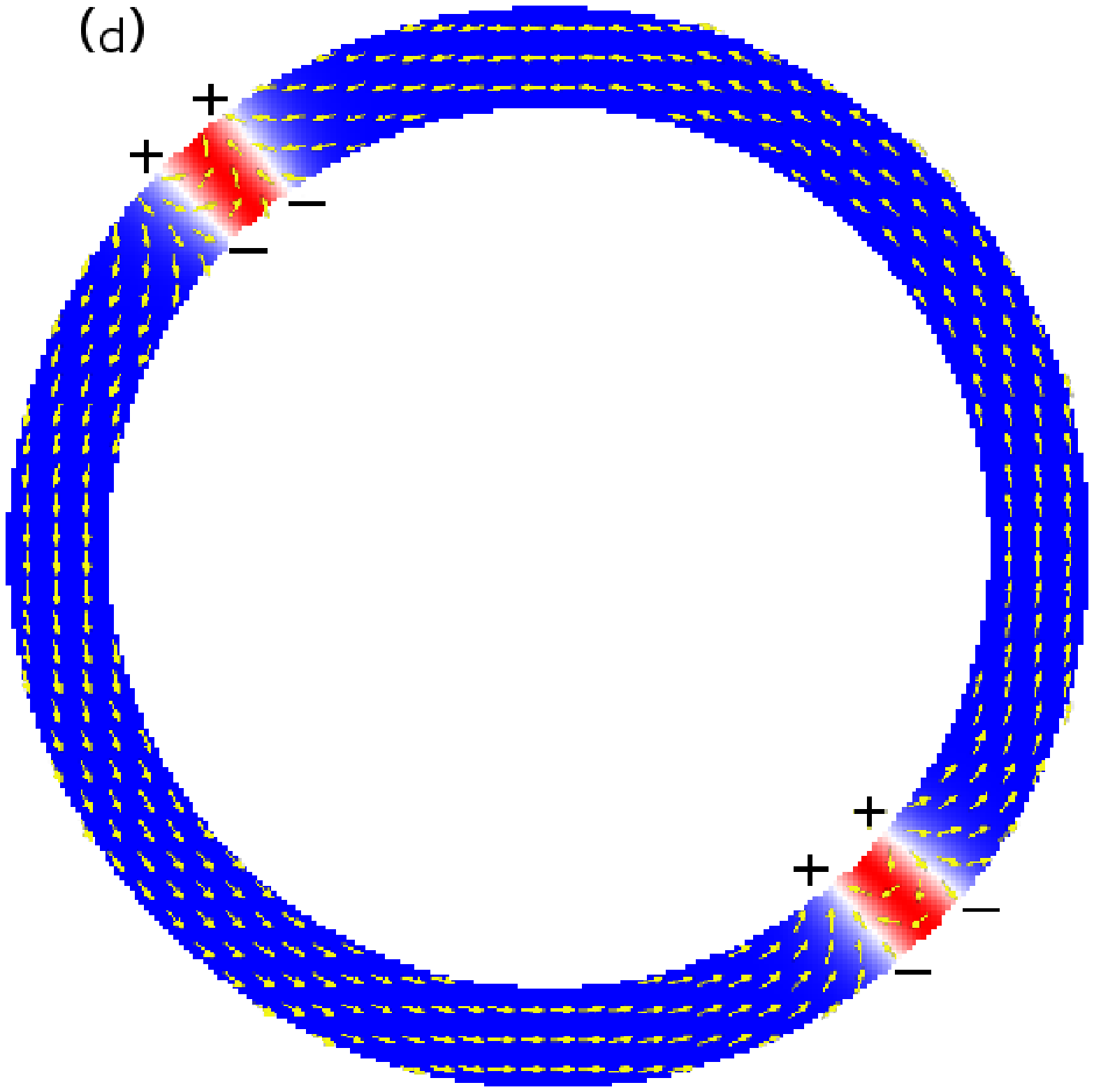}

\caption{\label{fig:highmetastable}a) Transverse domain wall composed of two
edge defects of opposite sign. b) Magnetic configuration equivalent
to the instanton state in a stripe. c) A trapped domain between two
antiparallel walls. d) Low energy metastable state configuration.
The topological defect sign at the inner boundary determines if the
domain wall pairs are stable with respect to the external magnetic
field. With respect to the ground state (counterclockwise), this state
has a winding number of zero, as do the instanton saddle, constant
saddle and clockwise configurations. }

\end{figure}

Fig.~\ref{fig:highmetastable}d shows one of these locally stable
double pair states. They can be described as a combination of topological
defects, in particular edge defects and domain walls, which we will
now describe (cf. Fig. \ref{fig:singularities}). Bulk topological
defects are vortex and antivortex singularities of the magnetization
configurations with a net contribution to the exchange energy\citep{Chaikin}.
They are characterized by their winding number which is conserved
over any continous transformation of the magnetization. Close to an
edge of the material the singularities become \textit{half-vortices}
with winding number $\pm1/2$\citep{Tchernyshyov}. Since $\ell\rightarrow\infty$
for the limit $m\rightarrow1$, we can consider any small segment
of the ring as a strip: the outer edge of the ring maps into the lower
part of the strip.

A transverse domain wall can be described as a composite of two edge
defects of opposite sign\citep{Tchernyshyov} at opposite sides of
a ferromagnetic strip (see Fig. \ref{fig:highmetastable}a). In the
ring, it is convenient to label such a domain wall using the sign
of the topological defect on the inner side of the ring. The topological
defects experience a {}``Coulomb-like'' attraction or repulsion.
Walls where the magnetization points in opposite directions (equal
signs for same side edge defects as in Fig. \ref{fig:highmetastable}c)
experience repulsive interactions. The origin of repulsion arises
from the magnetostatic and exchange energies in the region between
walls. Walls which are parallel to each other (with opposite signs
in the same side edge defects, as in Fig. \ref{fig:highmetastable}b)
experience attractive interactions.

Any small fluctuation of the magnetization initially parallel to the
ring (strip) edges would be a precursor to a double wall of this last
type (Fig.~\ref{fig:highmetastable}b) as illustrated by the profile
of such configurations. The radial component of the magnetization
in the ring (transversal component in the strip) edges could have
any sign: the fluctuation will have the same energy whether the magnetization
tilts towards the inside or the outside of the ring (up or down in
a strip). A reversal can be produced by a fluctuation with equal probability
for any of these two orientations.

The domain between two walls will expand under the influence of a
parallel external magnetic field (producing a repulsive pressure on
the walls), and contract under an antiparallel magnetic field (producing
an attractive pressure on the walls). The balance between the interdefect
interaction and the field determines whether a configuration is in
stable or unstable equilibrium. For example, an instanton saddle configuration
is equivalent to two domain walls with opposite signs on their innermost
defects which enclose a domain parallel to the field (cf. signs in
Fig.~\ref{fig:highmetastable}~b). The field pushes the domain walls
away from each other while their mutual interaction tends to bring
them together. These opposing tendencies produce the unstable equilibrium
which makes this configuration a saddle state.

The opposite situation, in which edge defects repel and the enclosed
domain is antiparallel to the field, produces a metastable state.
The enclosed magnetic domain does not vanish because the half vortices
experience an effective repulsion: it is energically costly to unwind
the topological defects. As a result this configuration is stable,
with an energy intermediate between the clockwise and counterclockwise
configurations.

\begin{figure}
\includegraphics[width=3in]{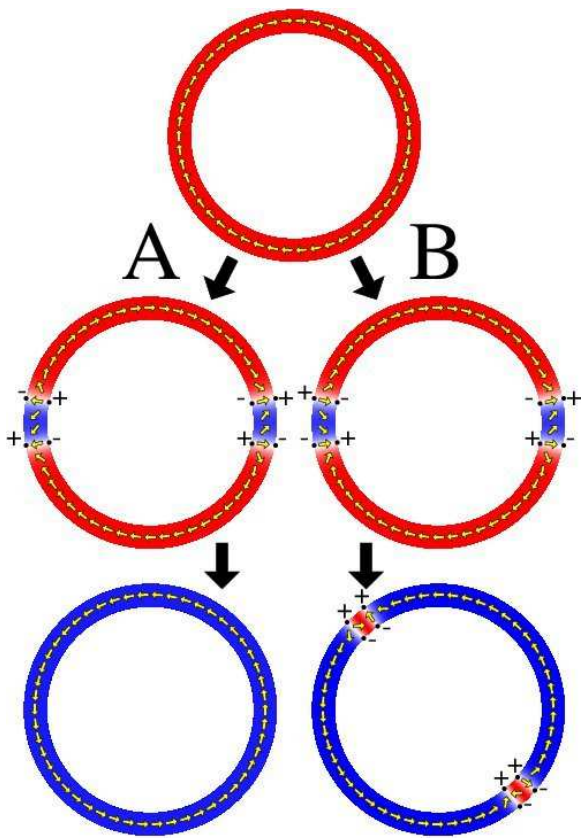}

\caption{\label{fig:appearanceofdoublewall}Two possible evolutions of a fluctuation
with two instantons in different parts of the ring. (A) The magnetization
relaxes to the metastable state only if the radial magnetization components
are parallel. (B) Otherwise, two trapped domains (360 degree walls)
appear.}

\end{figure}

We can now explain how the metastable state evolves into the state
represented in Fig.~\ref{fig:highmetastable}d. If the ring is originally
magnetized clockwise, the energy density per ring segment of instanton
fluctuation size is higher than the instanton energy. If the ring
is sufficiently large, two instantons (one domain wall pair each)
are produced (Fig. \ref{fig:appearanceofdoublewall}). If both instantons
point in the same radial direction the system evolves into the counterclockwise
state (Fig. \ref{fig:appearanceofdoublewall} A). However, if the
instanton fluctuations point in opposite radial directions (Fig. \ref{fig:appearanceofdoublewall}
B) the system evolves to the state shown with 4 domain walls .

The domain wall pairs need not be at opposite sides of the ring for
this configuration to be stable, as confirmed by displacing one of
the wall pairs by several angles and waiting for the system to relax.

\begin{figure}
\includegraphics[width=3in]{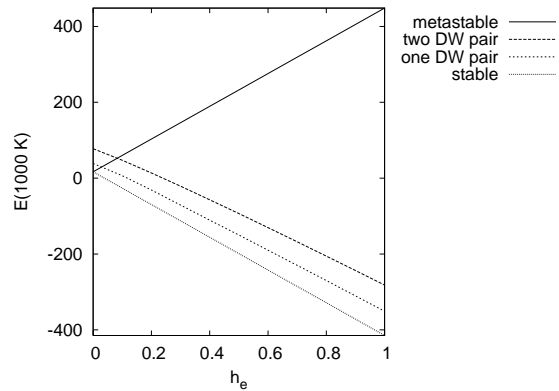}

\caption{\label{fig:Spectram1}Spectra of large ring regime $(m\approx1)$.
Lower metastable states are separated by the energy of a trapped domain
configuration. Beyond the clockwise configuration the instanton saddle
and the constant saddle are shown. The metastable configuration becomes
unstable beyond $h_{e}=1$.}

\end{figure}

It is interesting to summarize the new possible configurations and
their total energies in the large ring case. The lowest state is the
stable configuration, and there exists a series of metastable states
separated from each other by the energy of a double wall trapped domain
(Fig. \ref{fig:Spectram1})%

With the exception of the single wall pair, all configurations shown
have a total winding number difference from the counterclockwise state
equal to zero for a path that completely encloses the central hole
of the ring. The single wall pair configuration has this winding number
difference equal to one. While reversing the field will make the double
wall system decay into the stable state; the single wall pair configuration
cannot decay into the counterclockwise configuration. A trapped domain
configuration using a single pair of domain walls has already been
proposed for an MRAM device \citep{Moneck08}.

\section{Wide rings}

\label{sec:wide}

Having verified numerically the conclusions and predictions of the
analytical model of Martens\emph{~}\textit{\emph{et al.\citep{martens1}}}
for narrow rings, we now proceed to test the limits of its applicability
with increasing annular width. Given the $1D$ nature of the analytical
mode, we expect that at some width the model should break down and
its conclusions no longer apply. Surprisingly, this breakdown finally
occurs at a larger width than initially expected.

Increasing the ring width introduces two new effects that cause the
analytical model to break down. First, it allows the external field
to vary in magnitude appreciably as one moves along a radial direction.
Second, it increases the relative magnitude of the (previously neglected)
nonlocal bulk term of the magnetostatic energy with respect to that
of the local surface term.

In our simulations, $\Delta R$ is set to the values $100$ nm ($H_{c}=29.5$
mT), 200 nm ($H_{c}=14.7$ mT) and 380 nm ($H_{c}=$7.8 mT). The central
hole of the annulus is a few exchange lengths in diameter in the latter
case. For this reason vortex-like {}``singularities'' remain in
the gap.

\begin{figure}
\includegraphics[width=2in,angle=-90]{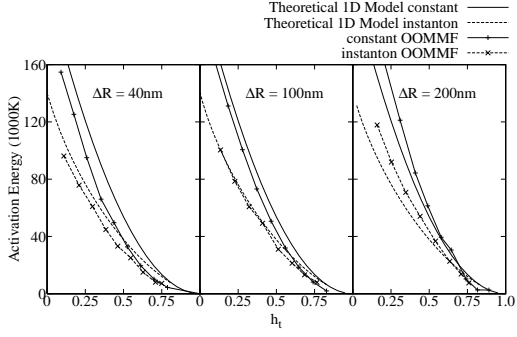}

\caption{\label{fig:changedeltaR} Activation energy spectra for $R=200$ nm,
$\ell=$12 and a) $\Delta R=40$ nm, b) 100 nm, c) 200 nm.}

\end{figure}

Fig.~\ref{fig:changedeltaR} summarizes the activation energies for
the widths considered. It should be noticed that as $\Delta R$ increases
for fixed $\lambda$, $\ell$ decreases and the annulus shifts away
from the large ring ($m\approx1$) approximation. We consequently
discuss only middle size rings where $\ell\approx12$. Fig.~\ref{fig:changedeltaR}
a,b,c shows that the main predictions of the model hold even for very
wide rings: the configurations $\phi_{h}$ are saddles for certain
$h_{t}$, and for fields below $h_{c}$ the instanton saddle configuration
is preferred to the constant saddle configuration.

\begin{figure}
\includegraphics[width=3in]{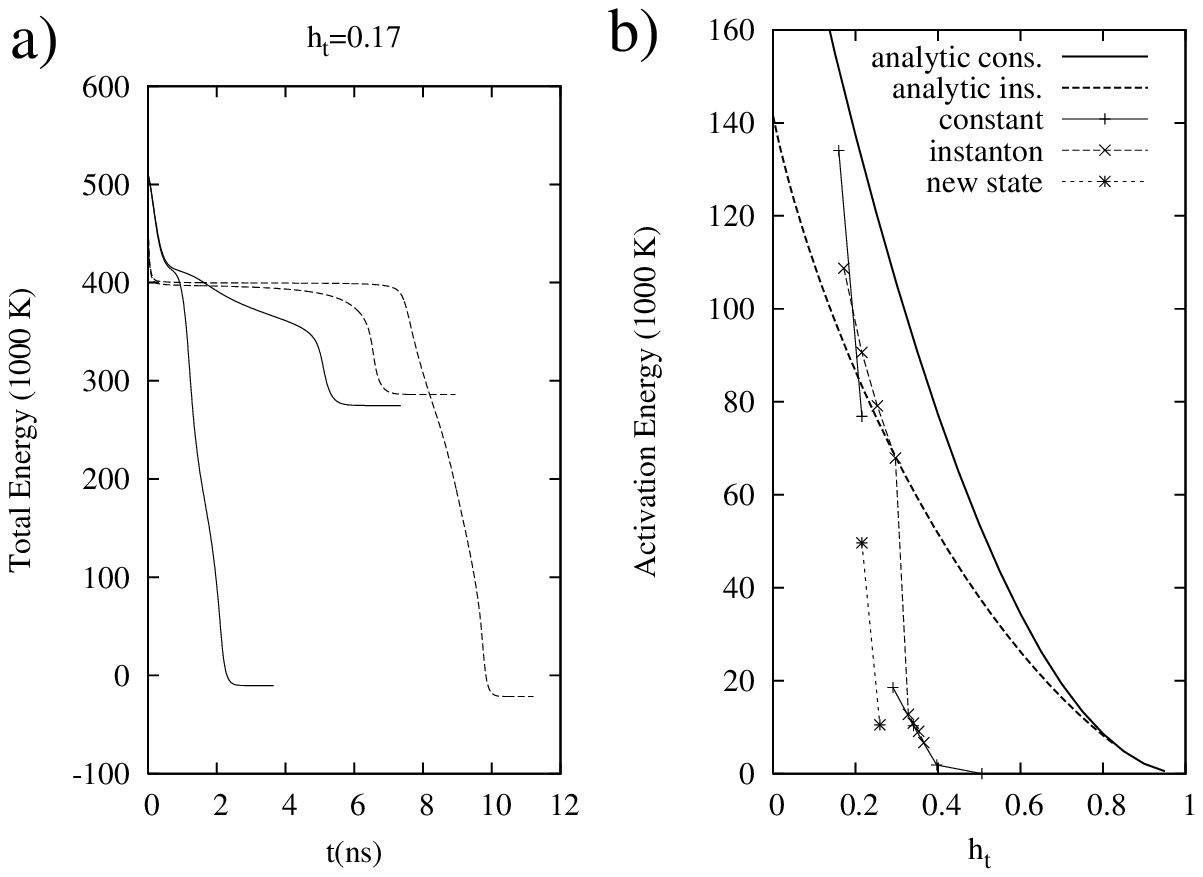}

\medskip{}

\includegraphics[width=1in]{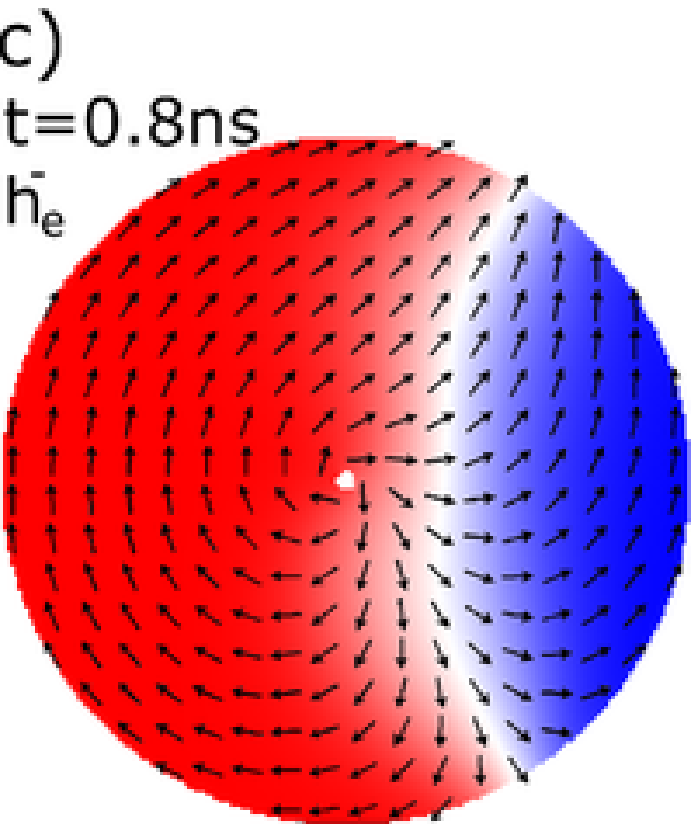}\includegraphics[width=1in]{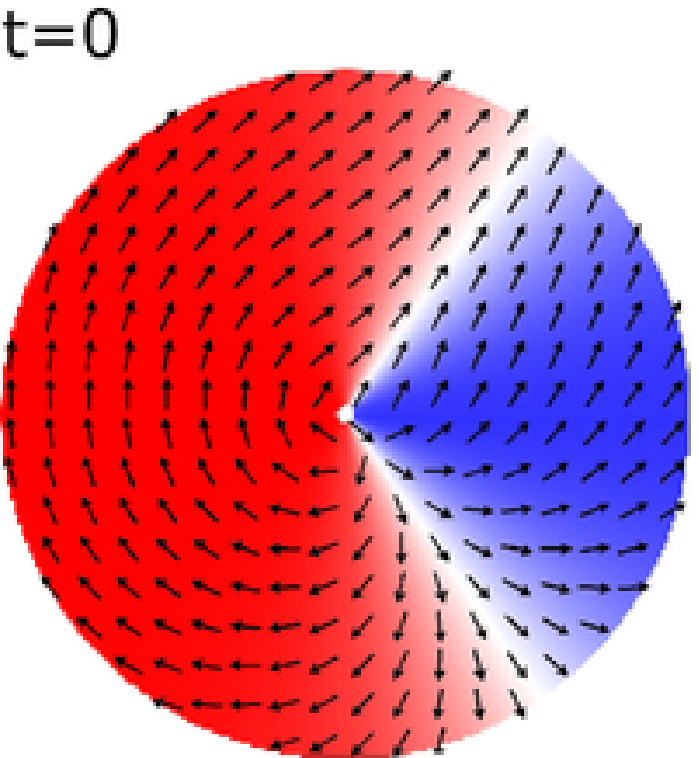}\includegraphics[width=1in]{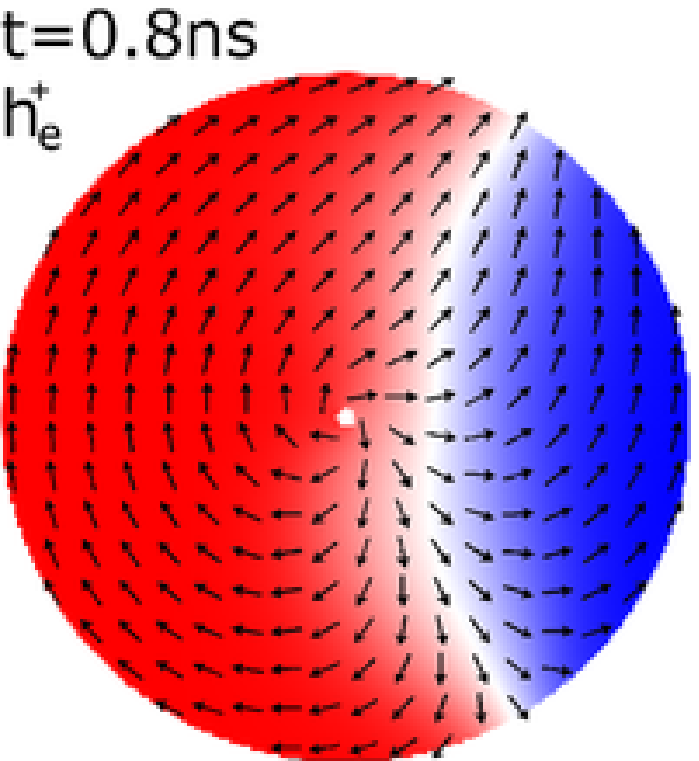}\medskip{}

\includegraphics[width=1in]{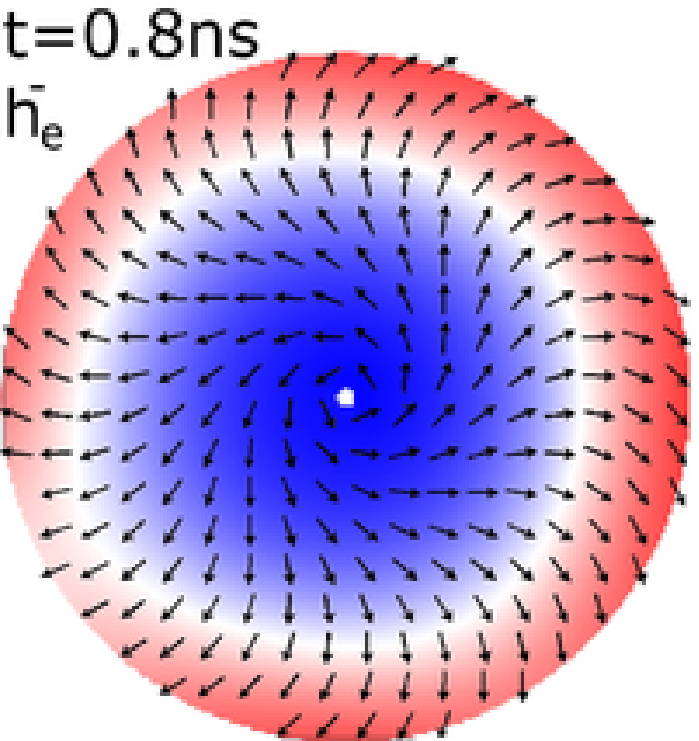}\includegraphics[width=1in]{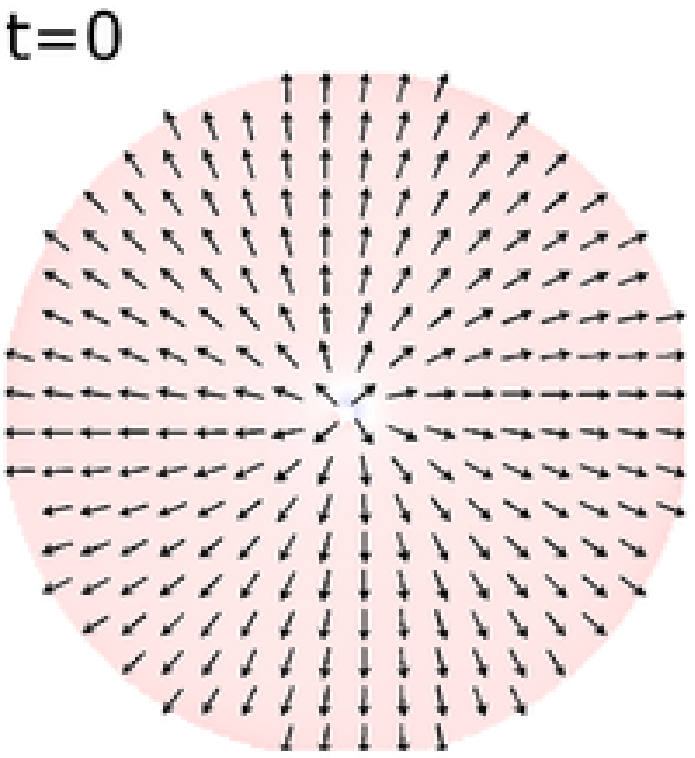}\includegraphics[width=1in]{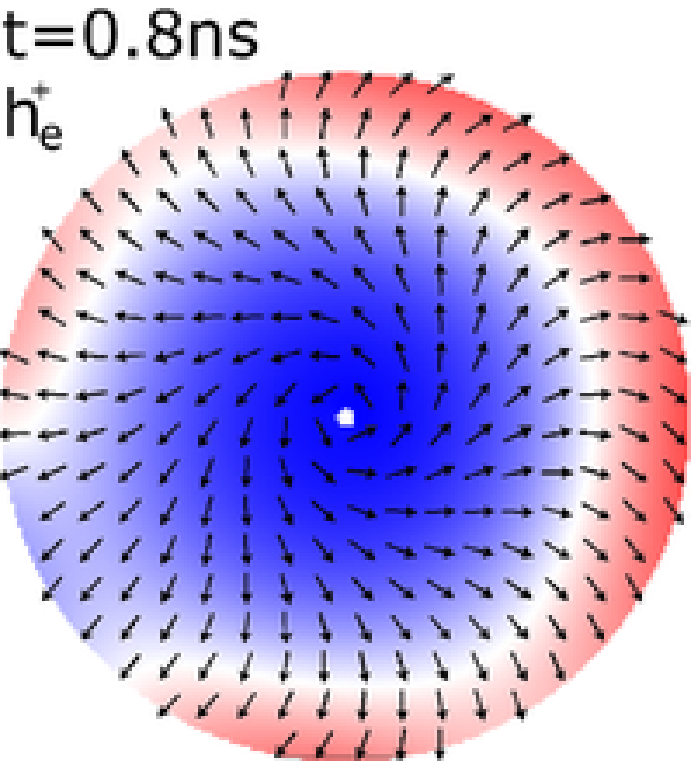}

\caption{\label{fig:failure}a) Time evolution of the total energy, for both
initial saddle configurations. At a given $h_{t}$ the dynamics do
not slow arbitrarily as $t\to0$. b) Spectra of the approximate activation
energies (measured close to the bifurcation in time, not at $t=0$).
c) The new metastable state of Fig.~\ref{fig:inhomogenousfield}
is found at lower energies than either saddle. The 1D saddle configurations
(instanton center-up, constant center-down) and their corresponding
relaxed states at $t$=0.8 ns for ($h_{e}^{+}$ right, $h_{e}^{-}$
left); inside the ring the strong field aligns the magnetization clockwise
.}

\end{figure}

We have found that the annular width must be increased to the extreme
wide-ring limit $\Delta R\approx2R$ in order for the model to fail.
Its breakdown can be observed in the $E(t)$ curve of Fig.~\ref{fig:failure}a.
In this regime, there are still fields $h_{t}$ for which the dynamics
bifurcates to either the stable or metastable state on either side
of $h_{t}$, but at such fields it is clear from Fig.~\ref{fig:failure}a
that $(\lim_{h\rightarrow h_{t}}\left.\frac{\delta E}{\delta t}\right|_{t=0}\ \not\rightarrow0^{-})$.
This indicates that the initial configurations chosen from the $1D$
analytical solutions are no longer close to the true saddles. Instead
of a long initial period of little change, we find instead relaxation
to a state in which the central region of the ring is magnetized circumferentially
and the outer edge of the ring retains some memory of the starting
configuration. This appears to be a new type of saddle configuration,
which we call the \textit{relaxed state\/}, and is shown in Fig.~\ref{fig:failure}.

For simulations starting from the instanton saddle with parameters
$h=0.1,...,0.4$ the system evolves to the relaxed state. At higher
values of $h$ ($h=0.5,...,0.8$ for the instanton, and $h=0.7,...,1.0$
for constant saddle configurations) the relaxed state does not satisfy
the stationarity condition, but the bifurcation condition can still
be satisfied with a particular $h_{t}$. The absence of a plateau
in the $E$ vs.~$t$ curves makes the definition of the activation
energy somewhat more problematic, but it can still be defined by using
as the energy of the saddle state the point at which the curves $E[\phi_{h},h_{e}^{-}]$
and $E[\phi_{h},h_{e}^{+}]$ separate. In this way approximate activation
energy spectra can be determined, and the results are shown in Fig.~\ref{fig:failure}b.

\begin{figure}
\includegraphics[width=2in]{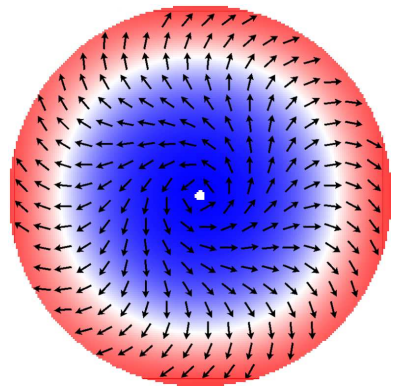}

\caption{\label{fig:inhomogenousfield} Magnetization configuration of new
found metastable state for extremely wide rings. The stability of
this state is a consequence of the highly inhomogeneous field in the
center.}

\end{figure}

For values of $h$ that correspond to the constant saddle (0.2,$\ldots$,0.6)
another feature of the breakdown of the $1D$ model can be seen. The
system relaxes to neither of the stable states considered so far (see
Fig.~\ref{fig:inhomogenousfield}). It evolves to a radially dependent
state with counterclockwise orientation at the inner edge of the ring
and clockwise orientation at the outer edge. This configuration is
stabilized by the large inhomogeneity in the magnitude of the magnetic
field as one moves outward along a radial direction. It is important
to note that the energy of this state is lower than the energies of
either initial configuration used (the 1D analytical saddle solutions),
but is higher than either the clockwise or counterclockwise state.

\begin{figure}
\includegraphics[width=2in,angle=-90]{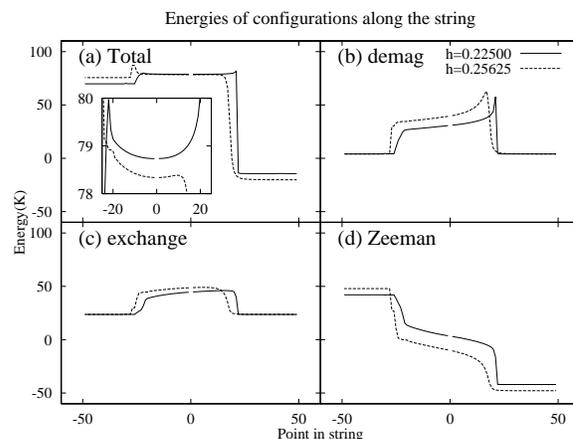}

\caption{\label{fig:relaxedstring}Energies of the configurations along the
string after 100 iterations at two different applied fields. The center
of the figure represents the magnetization configuration of Fig. \ref{fig:inhomogenousfield}.
The clockwise configuration is at -50, and the counterclockwise is
located at +50. Energies are a) total b) demagnetization c) exchange
d) Zeeman . The inset on (a) shows the energy on a finer scale, illustrating
the existence of an energy barrier.}

\end{figure}

There exists a low barrier that prevents this configuration from relaxing
into either of the counterclockwise configurations in the limiting
fields of Fig. \ref{fig:failure}. We obtain an estimate of this energy
by using the String Method without reparametrization\citep{Weinan2002}
as described in Sect. \ref{method}. We start with sequences of 50
equally spaced configurations that connect each of the circumferencial
configurations to this newly found metastable state along an straight
line. The result of the relaxations are shown in Fig. \ref{fig:relaxedstring}.
Fig. \ref{fig:relaxedstring} (a) shows the total energy of the string
points after relaxation; the inset shows a very shallow energy barrier
that prevents decay to the two lowest stable configurations. It is
interesting to observe that this barrier almost disappears on the
left at $h_{e}=0.22500$ and on the right at $h_{e}=0.25625$ explaining
why this state is stable only for a very narrow band of field values.

The physical origin of the two barriers is clear if the three components
are studied separately (Fig.~\ref{fig:relaxedstring}~b,~c,d).
For all graphs the origin corresponds to the state represented in
Fig.~\ref{fig:inhomogenousfield}, and the clockwise oriented magnetization
is 50 steps to the left of zero, the counterclockwise configuration
is located 50 steps to the right of zero.

The Zeeman energy (Fig.~\ref{fig:relaxedstring}d) prevents the system
from moving towards the clockwise configuration for all states along
the string. It is maximum for the clockwise configuration and decreases
monotonically along the path. Any point of the trajectory is pushed
to the right of the graph by the external field.

Fig.~\ref{fig:relaxedstring}b shows the demagnetization energy with
a sharp barrier that prevents the magnetization from pointing perpendicular
to the ring edges. At this barrier, the magnetization at the surface
points radially outward (as in Fig. \ref{fig:failure}$h_{e}^{+}$
for the constant saddle). This produces a sharp barrier at this step
of the path. At this point, the magnetostatic energy is the only energy
term acting against the reversal of the magnetization from clockwise
to counterclockwise orientations; other terms favors the reversal.
The net effect of the demagnetization energy barrier is to favor configurations
away from this barrier. Fig.~\ref{fig:relaxedstring}c represents
the exchange energy. The exchange energy is minimal in the two circumferencial
states.

When these three interactions are considered together the stability
of this state is understood: at lower fields the exchange and magnetostatic
energy are balanced by the Zeeman energy. At large fields Zeeman and
exchange favor a magnetization out of the ring's surface, when the
shape anisotropy barrier is overcome, the system reverses suddenly
into the counterclockwise configuration.

The analytical model presumed the field to be constant in the radial
direction as in the narrow ring case. Although the field inhomogeneity
eventually renders this assumption invalid, the $1D$ model is surprisingly
robust and breaks down only in the extreme case just considered.

\section{Conclusion}

The $1D$ analytical model of Martens~\textit{et al.\/}~\citep{martens1}
has been tested and confirmed using numerical simulations for a variety
of ring sizes and external magnetic fields fields that more closely
approximate realistic laboratory situations. Although it was initially
expected that the analytical model would apply only to narrow annuli,
our simulations show that it is surprisingly robust, eventually breaking
down only in the extreme two-dimensional limit.

By studying a large portion of the relevant parameter space, ($\lambda,\Delta R$),
we have also found new saddle and stable states. These findings enrich
our understanding of the energy landscape of ferromagnetic rings.

Two limits present particularly interesting features: the large-$R$
narrow ring ($\Delta R\ll R$) and the extremely wide ring. The large
narrow ring allows for the appearance of multiple instantons at energies
below the constant saddle configuration; their relative orientations
and positions determine the final magnetization configuration. We
provide a topological analysis of these new configurations and predict
a hierarchy of such states differing by the number of domain wall
pairs in each. The interaction between these domain walls can be characterized
and understood through {}``edge defects'' that compose them. Moreover,
by following the evolution of the downhill energy run from one of
these states to the stable configuration, one can infer the propagation
of domain wall pairs in the ring: sudden changes in energy indicate
annihilation of domain wall pairs, while linear decrease of energy
occurs during domain wall propagation.

The $1D$ model predicts extremely well the activation energy for
magnetization reversal even for wide rings. Eventually, though, in
the extremely wide regime limit, the $2D$ nature of the magnetic
field becomes important and the $1D$ approximation breaks down. In
this regime, a new stable state arises in which the magnetization
is radially dependent but independent of the angle. The String Method~\citep{Weinan2002}
has been used to verify the existence of a barrier between this state
and other stable states with lower energy, and is used to clarify
the physical origin of this barrier, in terms of the various contributions
to the energy from exchange, magnetostatic, and external field sources.

\begin{acknowledgments}
We want to thank Hans Fangohr for providing assistance in the use
of the Nmag simulation package. We also acknowledge Daniel Bedau for
helpful insight. This research has been partially supported by NSF-DMR-0706322
and an NYU-Research Challenge Fund award (ADK) and by NSF-PHYS-0651077
(DLS). We are grateful to Oleg Tretiakov for commenting on the first
draft of this manuscript. 
\end{acknowledgments}
\bibliographystyle{apsper}
\bibliography{nanoringreversal}

\begin{thebibliography}{23}
\expandafter\ifx\csname natexlab\endcsname\relax\def\natexlab#1{#1}\fi
\expandafter\ifx\csname bibnamefont\endcsname\relax
  \def\bibnamefont#1{#1}\fi
\expandafter\ifx\csname bibfnamefont\endcsname\relax
  \def\bibfnamefont#1{#1}\fi
\expandafter\ifx\csname citenamefont\endcsname\relax
  \def\citenamefont#1{#1}\fi
\expandafter\ifx\csname url\endcsname\relax
  \def\url#1{\texttt{#1}}\fi
\expandafter\ifx\csname urlprefix\endcsname\relax\def\urlprefix{URL }\fi
\providecommand{\bibinfo}[2]{#2}
\providecommand{\eprint}[2][]{\url{#2}}

\bibitem[{\citenamefont{Chien et~al.}(2007)\citenamefont{Chien, Zhu, and
  J.G.Zhu}}]{_patterned_2007}
\bibinfo{author}{\bibfnamefont{C.~L.} \bibnamefont{Chien}},
  \bibinfo{author}{\bibfnamefont{F.}~\bibnamefont{Zhu}}, \bibnamefont{and}
  \bibinfo{author}{\bibnamefont{J.G.Zhu}}, \bibinfo{journal}{Phys. Today}
  \textbf{\bibinfo{volume}{60}}, \bibinfo{pages}{40} (\bibinfo{year}{2007}).

\bibitem[{\citenamefont{Yang et~al.}(2007)\citenamefont{Yang, Hara, Hirohata,
  Kimura, and Otani}}]{yang2007}
\bibinfo{author}{\bibfnamefont{T.}~\bibnamefont{Yang}},
  \bibinfo{author}{\bibfnamefont{M.}~\bibnamefont{Hara}},
  \bibinfo{author}{\bibfnamefont{A.}~\bibnamefont{Hirohata}},
  \bibinfo{author}{\bibfnamefont{T.}~\bibnamefont{Kimura}}, \bibnamefont{and}
  \bibinfo{author}{\bibfnamefont{Y.}~\bibnamefont{Otani}},
  \bibinfo{journal}{Appl. Phys. Lett.} \textbf{\bibinfo{volume}{90}},
  \bibinfo{pages}{022504} (\bibinfo{year}{2007}).

\bibitem[{\citenamefont{Zhu et~al.}(2000)\citenamefont{Zhu, Zheng, and
  Prinz}}]{zhu2000}
\bibinfo{author}{\bibfnamefont{J.-G.} \bibnamefont{Zhu}},
  \bibinfo{author}{\bibfnamefont{Y.}~\bibnamefont{Zheng}}, \bibnamefont{and}
  \bibinfo{author}{\bibfnamefont{G.~A.} \bibnamefont{Prinz}}, in
  \emph{\bibinfo{booktitle}{J. Appl. Phys.}} (\bibinfo{publisher}{AIP},
  \bibinfo{year}{2000}), vol.~\bibinfo{volume}{87}, pp.
  \bibinfo{pages}{6668--6673}.

\bibitem[{\citenamefont{Kl\"aui et~al.}(2003)\citenamefont{Kl\"aui, Vaz, Bland,
  Monchesky, Unguris, Bauer, Cherifi, Heun, Locatelli, Heyderman
  et~al.}}]{Klaui2003}
\bibinfo{author}{\bibfnamefont{M.}~\bibnamefont{Kl\"aui}},
  \bibinfo{author}{\bibnamefont{Vaz}}, \bibinfo{author}{\bibfnamefont{J.~A.~C.}
  \bibnamefont{Bland}}, \bibinfo{author}{\bibfnamefont{T.~L.}
  \bibnamefont{Monchesky}},
  \bibinfo{author}{\bibfnamefont{J.}~\bibnamefont{Unguris}},
  \bibinfo{author}{\bibfnamefont{E.}~\bibnamefont{Bauer}},
  \bibinfo{author}{\bibfnamefont{S.}~\bibnamefont{Cherifi}},
  \bibinfo{author}{\bibfnamefont{S.}~\bibnamefont{Heun}},
  \bibinfo{author}{\bibfnamefont{A.}~\bibnamefont{Locatelli}},
  \bibinfo{author}{\bibfnamefont{L.~J.} \bibnamefont{Heyderman}},
  \bibnamefont{et~al.}, \bibinfo{journal}{Phys. Rev. B}
  \textbf{\bibinfo{volume}{68}}, \bibinfo{pages}{134426}
  (\bibinfo{year}{2003}).

\bibitem[{\citenamefont{Kl\"aui et~al.}(2004)\citenamefont{Kl\"aui, Vaz, Bland,
  Heyderman, Nolting, Pavlovska, Bauer, Cherifi, Heun, and
  Locatelli}}]{klaui2004}
\bibinfo{author}{\bibfnamefont{M.}~\bibnamefont{Kl\"aui}},
  \bibinfo{author}{\bibfnamefont{C.~A.~F.} \bibnamefont{Vaz}},
  \bibinfo{author}{\bibfnamefont{J.~A.~C.} \bibnamefont{Bland}},
  \bibinfo{author}{\bibfnamefont{L.~J.} \bibnamefont{Heyderman}},
  \bibinfo{author}{\bibfnamefont{F.}~\bibnamefont{Nolting}},
  \bibinfo{author}{\bibfnamefont{A.}~\bibnamefont{Pavlovska}},
  \bibinfo{author}{\bibfnamefont{E.}~\bibnamefont{Bauer}},
  \bibinfo{author}{\bibfnamefont{S.}~\bibnamefont{Cherifi}},
  \bibinfo{author}{\bibfnamefont{S.}~\bibnamefont{Heun}}, \bibnamefont{and}
  \bibinfo{author}{\bibfnamefont{A.}~\bibnamefont{Locatelli}},
  \bibinfo{journal}{Appl. Phys. Lett.} \textbf{\bibinfo{volume}{85}},
  \bibinfo{pages}{5637} (\bibinfo{year}{2004}).

\bibitem[{\citenamefont{Laufenberg et~al.}(2006)\citenamefont{Laufenberg,
  Backes, Buhrer, Bedau, Klaui, Rudiger, Vaz, Bland, Heyderman, Nolting
  et~al.}}]{Laufenberg}
\bibinfo{author}{\bibfnamefont{M.}~\bibnamefont{Laufenberg}},
  \bibinfo{author}{\bibfnamefont{D.}~\bibnamefont{Backes}},
  \bibinfo{author}{\bibfnamefont{W.}~\bibnamefont{Buhrer}},
  \bibinfo{author}{\bibfnamefont{D.}~\bibnamefont{Bedau}},
  \bibinfo{author}{\bibfnamefont{M.}~\bibnamefont{Klaui}},
  \bibinfo{author}{\bibfnamefont{U.}~\bibnamefont{Rudiger}},
  \bibinfo{author}{\bibfnamefont{C.~A.~F.} \bibnamefont{Vaz}},
  \bibinfo{author}{\bibfnamefont{J.~A.~C.} \bibnamefont{Bland}},
  \bibinfo{author}{\bibfnamefont{L.~J.} \bibnamefont{Heyderman}},
  \bibinfo{author}{\bibfnamefont{F.}~\bibnamefont{Nolting}},
  \bibnamefont{et~al.}, \bibinfo{journal}{Appl. Phys. Lett.}
  \textbf{\bibinfo{volume}{88}}, \bibinfo{pages}{052507}
  (\bibinfo{year}{2006}).

\bibitem[{\citenamefont{{Casta\~no} et~al.}(2003)\citenamefont{{Casta\~no},
  Ross, Frandsen, Eilez, Gil, Smith, Redjdal, and Humphrey}}]{castano2003}
\bibinfo{author}{\bibfnamefont{F.~J.} \bibnamefont{{Casta\~no}}},
  \bibinfo{author}{\bibfnamefont{C.~A.} \bibnamefont{Ross}},
  \bibinfo{author}{\bibfnamefont{C.}~\bibnamefont{Frandsen}},
  \bibinfo{author}{\bibfnamefont{A.}~\bibnamefont{Eilez}},
  \bibinfo{author}{\bibfnamefont{D.}~\bibnamefont{Gil}},
  \bibinfo{author}{\bibfnamefont{H.~I.} \bibnamefont{Smith}},
  \bibinfo{author}{\bibfnamefont{M.}~\bibnamefont{Redjdal}}, \bibnamefont{and}
  \bibinfo{author}{\bibfnamefont{F.~B.} \bibnamefont{Humphrey}},
  \bibinfo{journal}{Phys. Rev. B} \textbf{\bibinfo{volume}{67}},
  \bibinfo{pages}{184425} (\bibinfo{year}{2003}).

\bibitem[{\citenamefont{{Casta\~no} et~al.}(2006)\citenamefont{{Casta\~no},
  Morecroft, and Ross}}]{castano2006}
\bibinfo{author}{\bibfnamefont{F.~J.} \bibnamefont{{Casta\~no}}},
  \bibinfo{author}{\bibfnamefont{D.}~\bibnamefont{Morecroft}},
  \bibnamefont{and} \bibinfo{author}{\bibfnamefont{C.~A.} \bibnamefont{Ross}},
  \bibinfo{journal}{Physical Rev. B (Condensed Matter and Materials Phys.)}
  \textbf{\bibinfo{volume}{74}}, \bibinfo{pages}{224401}
  (\bibinfo{year}{2006}).

\bibitem[{\citenamefont{Rothman et~al.}(2001)\citenamefont{Rothman, Kl�ui,
  Lopez-Diaz, Vaz, Bleloch, Bland, Cui, and Speaks}}]{rothman2001}
\bibinfo{author}{\bibfnamefont{J.}~\bibnamefont{Rothman}},
  \bibinfo{author}{\bibfnamefont{M.}~\bibnamefont{Kl�ui}},
  \bibinfo{author}{\bibfnamefont{L.}~\bibnamefont{Lopez-Diaz}},
  \bibinfo{author}{\bibfnamefont{C.~A.~F.} \bibnamefont{Vaz}},
  \bibinfo{author}{\bibfnamefont{A.}~\bibnamefont{Bleloch}},
  \bibinfo{author}{\bibfnamefont{J.~A.~C.} \bibnamefont{Bland}},
  \bibinfo{author}{\bibfnamefont{Z.}~\bibnamefont{Cui}}, \bibnamefont{and}
  \bibinfo{author}{\bibfnamefont{R.}~\bibnamefont{Speaks}},
  \bibinfo{journal}{Phys. Rev. Lett.} \textbf{\bibinfo{volume}{86}},
  \bibinfo{pages}{1098} (\bibinfo{year}{2001}).

\bibitem[{\citenamefont{Martens et~al.}(2006)\citenamefont{Martens, Stein, and
  Kent}}]{martens1}
\bibinfo{author}{\bibfnamefont{K.}~\bibnamefont{Martens}},
  \bibinfo{author}{\bibfnamefont{D.~L.} \bibnamefont{Stein}}, \bibnamefont{and}
  \bibinfo{author}{\bibfnamefont{A.~D.} \bibnamefont{Kent}},
  \bibinfo{journal}{Phys. Rev. B (Condensed Matter and Materials Phys.)}
  \textbf{\bibinfo{volume}{73}}, \bibinfo{pages}{054413}
  (\bibinfo{year}{2006}).

\bibitem[{\citenamefont{Chaves-O'Flynn
  et~al.}(2008)\citenamefont{Chaves-O'Flynn, Xiao, Stein, and
  Kent}}]{chaves2008}
\bibinfo{author}{\bibfnamefont{G.~D.} \bibnamefont{Chaves-O'Flynn}},
  \bibinfo{author}{\bibfnamefont{K.}~\bibnamefont{Xiao}},
  \bibinfo{author}{\bibfnamefont{D.~L.} \bibnamefont{Stein}}, \bibnamefont{and}
  \bibinfo{author}{\bibfnamefont{A.~D.} \bibnamefont{Kent}},
  \bibinfo{journal}{J. Appl. Phys.} \textbf{\bibinfo{volume}{103}},
  \bibinfo{pages}{07D917} (\bibinfo{year}{2008}).

\bibitem[{\citenamefont{Kramers}(1940)}]{Kramers40}
\bibinfo{author}{\bibfnamefont{H.~A.} \bibnamefont{Kramers}},
  \bibinfo{journal}{Physica} \textbf{\bibinfo{volume}{7}}, \bibinfo{pages}{284}
  (\bibinfo{year}{1940}).

\bibitem[{\citenamefont{H\"anggi et~al.}(1990)\citenamefont{H\"anggi, Talkner,
  and Borkovec}}]{Hangii90}
\bibinfo{author}{\bibfnamefont{P.}~\bibnamefont{H\"anggi}},
  \bibinfo{author}{\bibfnamefont{P.}~\bibnamefont{Talkner}}, \bibnamefont{and}
  \bibinfo{author}{\bibfnamefont{M.}~\bibnamefont{Borkovec}},
  \bibinfo{journal}{rev. of Modern Phys.} \textbf{\bibinfo{volume}{62}},
  \bibinfo{pages}{251} (\bibinfo{year}{1990}).

\bibitem[{\citenamefont{Gilbert}(1955)}]{Gilbert55}
\bibinfo{author}{\bibfnamefont{T.~L.} \bibnamefont{Gilbert}},
  \bibinfo{journal}{Phys. Rev.} \textbf{\bibinfo{volume}{100}},
  \bibinfo{pages}{1243} (\bibinfo{year}{1955}).

\bibitem[{\citenamefont{Landau and Lifshitz}(1935)}]{Landau35}
\bibinfo{author}{\bibfnamefont{L.}~\bibnamefont{Landau}} \bibnamefont{and}
  \bibinfo{author}{\bibfnamefont{E.}~\bibnamefont{Lifshitz}},
  \bibinfo{journal}{Physik. Z} \textbf{\bibinfo{volume}{8}},
  \bibinfo{pages}{152} (\bibinfo{year}{1935}).

\bibitem[{\citenamefont{Abramowitz and Stegun}(1965)}]{abramowitzstegun}
\bibinfo{author}{\bibfnamefont{M.}~\bibnamefont{Abramowitz}} \bibnamefont{and}
  \bibinfo{author}{\bibfnamefont{I.~A.} \bibnamefont{Stegun}}
  (\bibinfo{publisher}{Dover Publications}, \bibinfo{year}{1965}), ISBN
  \bibinfo{isbn}{0486612724}.

\bibitem[{\citenamefont{Donahue and Porter}(1999)}]{donahue1999}
\bibinfo{author}{\bibfnamefont{M.}~\bibnamefont{Donahue}} \bibnamefont{and}
  \bibinfo{author}{\bibfnamefont{D.}~\bibnamefont{Porter}},
  \bibinfo{organization}{National Institute of Standards and Technology},
  \bibinfo{address}{Gaithersburg, MD}, \bibinfo{edition}{version 1.0} ed.
  (\bibinfo{year}{1999}).

\bibitem[{\citenamefont{Fischbacher et~al.}(2007)\citenamefont{Fischbacher,
  Franchin, Bordignon, and Fangohr}}]{Fischbacher2007}
\bibinfo{author}{\bibfnamefont{T.}~\bibnamefont{Fischbacher}},
  \bibinfo{author}{\bibfnamefont{M.}~\bibnamefont{Franchin}},
  \bibinfo{author}{\bibfnamefont{G.}~\bibnamefont{Bordignon}},
  \bibnamefont{and} \bibinfo{author}{\bibfnamefont{H.}~\bibnamefont{Fangohr}},
  \bibinfo{journal}{Magnetics, IEEE Transactions on}
  \textbf{\bibinfo{volume}{43}}, \bibinfo{pages}{2896} (\bibinfo{year}{2007}).

\bibitem[{\citenamefont{E et~al.}(2002)\citenamefont{E, Ren, and
  Vanden-Eijnden}}]{Weinan2002}
\bibinfo{author}{\bibfnamefont{W.}~\bibnamefont{E}},
  \bibinfo{author}{\bibfnamefont{W.}~\bibnamefont{Ren}}, \bibnamefont{and}
  \bibinfo{author}{\bibfnamefont{E.}~\bibnamefont{Vanden-Eijnden}},
  \bibinfo{journal}{Phys. Rev. B} \textbf{\bibinfo{volume}{66}},
  \bibinfo{pages}{052301} (\bibinfo{year}{2002}).

\bibitem[{\citenamefont{Braun}(1993)}]{Braun}
\bibinfo{author}{\bibfnamefont{H.-B.} \bibnamefont{Braun}},
  \bibinfo{journal}{Phys. Rev. Lett.} \textbf{\bibinfo{volume}{71}},
  \bibinfo{pages}{3557} (\bibinfo{year}{1993}).

\bibitem[{\citenamefont{Chaikin and Lubensky}(2000)}]{Chaikin}
\bibinfo{author}{\bibfnamefont{P.~M.} \bibnamefont{Chaikin}} \bibnamefont{and}
  \bibinfo{author}{\bibfnamefont{T.~C.} \bibnamefont{Lubensky}}
  (\bibinfo{publisher}{Cambridge University Press}, \bibinfo{year}{2000}),
  \bibinfo{edition}{1st} ed., ISBN \bibinfo{isbn}{0521794501}.

\bibitem[{\citenamefont{Tchernyshyov and Chern}(2005)}]{Tchernyshyov}
\bibinfo{author}{\bibfnamefont{O.}~\bibnamefont{Tchernyshyov}}
  \bibnamefont{and} \bibinfo{author}{\bibfnamefont{G.-W.} \bibnamefont{Chern}},
  \bibinfo{journal}{Phys. Rev. Lett.} \textbf{\bibinfo{volume}{95}},
  \bibinfo{pages}{197204} (\bibinfo{year}{2005}).

\bibitem[{\citenamefont{Moneck and J.Zhu}(2007)}]{Moneck08}
\bibinfo{author}{\bibfnamefont{M.}~\bibnamefont{Moneck}} \bibnamefont{and}
  \bibinfo{author}{\bibnamefont{J.Zhu}} (\bibinfo{year}{2007}), no.
  \bibinfo{number}{52nd} in \bibinfo{series}{Annual Conference on Magnetism and
  Magnetic Materials, MMM}.

\end{thebibliography}
\end{document}